\documentclass[10pt,twocolumn,twoside]{IEEEtran}
\usepackage{epsfig}
\usepackage{latexsym}
\usepackage{amssymb,amsfonts,amsmath,fmtcount,stmaryrd,mathrsfs,verbatim}
\usepackage[caption=false,font=footnotesize]{subfig}

\def\QQ{{\rlap {\raise 0.4ex \hbox{$\scriptscriptstyle |$}}\hskip -0.2em Q}}

\begin{document}

\title{An Information Theoretic Algorithm for Finding Periodicities in Stellar Light Curves}

\author{Pablo Huijse, \IEEEmembership{Student Member, IEEE}, Pablo A. Est\'evez*, \IEEEmembership{Senior Member, IEEE}, Pavlos Protopapas, Pablo Zegers, \IEEEmembership{Senior Member, IEEE}, and Jos\'e C. Pr\'incipe, \IEEEmembership{Fellow Member, IEEE} \thanks{Copyright (c) 2012 IEEE. Personal use of this material is permitted. However, permission to use this material for any other purposes must be obtained from the IEEE by sending a request to pubs-permissions@ieee.org.} \thanks{Pablo Huijse and Pablo Est\'evez* are with the Department of Electrical Engineering and the Advanced Mining Technology Center, Faculty of Physical and Mathematical Sciences, Universidad de Chile, Chile. *P. A. Est\'evez is the corresponding author.  Pavlos Protopapas is with the School of Engineering and Applied Sciences and the Center of Astrophysics, Harvard University, USA. Pablo Zegers is with the Autonomous Machines Center of the College of Engineering and Applied Sciences, Universidad de los Andes, Chile. Jose Principe is with the Computational Neuroengineering Laboratory, University of Florida, Gainsville, USA. Correspondent email address: \textit{pestevez@ing.uchile.cl}.}}

\markboth{IEEE Transactions on Signal Processing,~Vol.~1, No.~1, January~2012}{HUIJSE \MakeLowercase{\textit{et al.}}: An Information Theoretic Algorithm for Finding Periodicities in Stellar Light Curves}

\maketitle

\begin{abstract}
We propose a new information theoretic metric for finding periodicities in stellar light curves. Light curves are astronomical time series of brightness over time, and are characterized as being noisy and unevenly sampled. The proposed metric combines correntropy (generalized correlation) with a periodic kernel to measure similarity among samples separated by a given period. The new metric provides a periodogram, called Correntropy Kernelized Periodogram (CKP), whose peaks are associated with the fundamental frequencies present in the data. The CKP does not require any resampling, slotting or folding scheme as it is computed directly from the available samples. CKP is the main part of a fully-automated pipeline for periodic light curve discrimination to be used in astronomical survey databases. We show that the CKP method outperformed the slotted correntropy, and conventional methods used in astronomy for periodicity discrimination and period estimation tasks, using a set of light curves drawn from the MACHO survey. The proposed metric achieved 97.2\% of true positives with 0\% of false positives at the confidence level of 99\% for the periodicity discrimination task; and 88\% of hits with 11.6\% of multiples and 0.4\% of misses in the period estimation task.


\end{abstract}

\begin{IEEEkeywords}
Correntropy, information theory, time series analysis, period detection, period estimation, variable stars
\end{IEEEkeywords}

\IEEEpeerreviewmaketitle

\section{Introduction}

A light curve represents the brightness of a celestial object as a function of time (usually the magnitude of the star in the visible part of the electromagnetic radiation). Light curve analysis is an important tool in astrophysics used for estimation of stellar masses and distances to astronomical objects. By analyzing the light curves derived from the sky surveys, astronomers can perform tasks such as transient event detection, variable star detection and classification.



There are a certain types of variable stars \cite{Petit1997} whose brightness varies following regular cycles. Examples of this kind of stars are the pulsating variables and eclipsing binary stars. Pulsating stars, such as Cepheids and RR Lyrae, expand and contract periodically effectively changing their size, temperature and brightness. Eclipsing binaries, are systems of two stars with a common center of mass whose orbital plane is aligned to Earth. Periodic drops in brightness are observed due to the mutual eclipses between the components of the system. Although most stars have at least some variation in luminosity, current ground based survey estimations indicate that 3\% of the stars varying more than the sensitivity of the instruments and $\sim$1\% are periodic \cite{Eyer1999}.

Detecting periodicity and estimating the period of stars is of high importance in astronomy. The period is a key feature for classifying variable stars \cite{Debosscher2007, Wachman2009}; and estimating other parameters such as mass and distance to Earth \cite{Popper1980}. Period finding in light curves is also used as a means to find extrasolar planets \cite{protopapas2005}.
Light curve analysis is a particularly challenging task. Astronomical time series are unevenly sampled due to constraints in the observation schedules, the day-night cycle, bad weather conditions, equipment positioning, calibration and maintenance. Light curves are also affected by several noise sources such as light contamination from other astronomical sources near the line of sight (sky background), the atmosphere, the instruments and particularly the CCD detectors, among others. Moreover, spurious periods of one day, one month and one year are usually present in the data due to changes in the atmosphere and moon brightness.
Another challenge of light curve analysis is related to the number and size of the databases being build by astronomical surveys. Each observation phase of astronomical surveys such as  MACHO \cite{Alcock2000}, OGLE \cite{Udalski1997}, SDSS \cite{York2000} and Pan-STARRS \cite{Kaiser2002} have captured tens of millions of light curves. Soon to arrive survey projects such as the LSST \cite{LSST2012}  will collect approximately 30 terabytes of data per night which translates into databases of 10 billion stars.

Currently, most periodicity finding schemes used in astronomy are interactive and/or rely somehow on human visual inspection. This calls for automated and efficient computational methods capable of performing robust light curve analysis for large astronomical databases.

In this paper we propose the Correntropy Kernelized Periodogram (CKP), a new metric for finding periodicities. The CKP combines the information theoretic learning (ITL) concept of correntropy \cite{Principe2010} with a periodic kernel. The proposed metric yields a periodogram whose peaks are associated with the fundamental frequencies present in the data. A statistical criterion, based on the CKP, is used for periodicity discrimination. We demonstrate the advantages of using the CKP by comparing it with conventional spectrum estimators and related methods in databases of astronomical light curves.

\section{Related methods}
Several methods have been developed to cope with the characteristics of light curves. The most widely used are the Lomb-Scargle (LS) periodogram \cite{Lomb1976,Scargle1982}, epoch folding, analysis of variance (AoV) \cite{Schwarzenberg1989}, string length (SL) methods \cite{Dworetsky1983,Clarke2002}, and the discrete or slotted autocorrelation \cite{Mayo1974,Edelson1988}. For the LS and AoV periodograms statistical confidence measures have been developed to assess periodicity detection besides estimating the period.

The LS periodogram is an extension of the conventional periodogram for unevenly sampled time series. A sample power spectrum is obtained by fitting a trigonometric model in a least squares sense over the available randomly sampled data points. The maximum of the LS power spectrum corresponds to the angular frequency whose model best fits the time series. In epoch folding a trial period $P_t$ is used to obtain a phase diagram of the light curve by applying the modulus (mod) transformation of the time axis:
\[
\phi_i(P_t)=\frac{t_i \text{ mod } P_t}{P_t},
\]
where $t_i$ are the time instants of the light curve. The trial period $P_t$ is found by and ad-hoc method, or simply corresponds to a sweep among a range of values. This transformation is equivalent to dividing the light curve in segments of length $P_t$ and then plotting the segments one on top of  another, hence folding the light curve. If the true period is used to fold the light curve, the periodic shape will be clearly seen in the phase diagram. If a wrong period is used instead, the phase diagram won't show a clear structure and it will look like noise.

In AoV \cite{Schwarzenberg1989} the folded light curve is binned and the ratio of the within-bins variance and the between-bins variance is computed. If the light curve is folded using its true period, the AoV statistic is expected to reach a minimum value. In SL methods, the light curve is folded using a trial period and the sum of distances  between consecutive points (string) in the folded curve is computed. The true period is estimated by minimizing the string length on a range of trial periods. The true period is expected to yield the most ordered folded curve and hence the minimum total distance between points. In slotted autocorrelation \cite{Mayo1974,Edelson1988}, time lags are defined as intervals or slots instead of single values. The slotted autocorrelation function at a certain time lag slot is computed by averaging the cross product between samples whose time differences fall in the given slot.

All the related methods described above are based on second-order statistic analysis. Information theoretic based criteria extract information from the probability density function, therefore it includes higher-order statistical moments present in the data. This usually implies a better modelling of the underlying process and robustness to noise and outliers.

The slotting technique was extended to the information theoretic concept of correntropy in \cite{Huijse2011}. The slotted correntropy estimator was compared with the other mentioned techniques on period estimation of light curves from the MACHO survey, performing equally well on Cepheid/RR Lyrae and much better in eclipsing binaries period estimation. However, the slotted technique has the drawback that is highly dependant on the slot size.

\section{Background on ITL methods and periodic kernels}

\subsection{Generalized correlation function: Correntropy}\label{sec:correntropy}

In \cite{Principe2010,Santamaria2006} an information theoretic functional capable of measuring the statistical magnitude distribution and the time structure of random processes was introduced. The generalized correlation function (GCF) or correntropy measures similarities between feature vectors separated by a certain time delay in input space. The similarities are measured in terms of inner products in a high-dimensional kernel space. For a random process $\{X_t,t\in T\}$ with T being an index set, the correntropy function is defined as

\begin{equation} \label{Corr1}
	V(t_1,t_2)=  {\mathbb E_{x_{t_1}x_{t_2}}} [\kappa(x_{t_1},x_{t_2})],
\end{equation}

and the centered correntropy is defined as

\begin{equation} \label{centeredCorr}
\begin{split}
	U(t_1,t_2)=  {\mathbb E_{x_{t_1}x_{t_2}}} [\kappa(x_{t_1},x_{t_2})] \\ -  {\mathbb E_{x_{t_1}}} {\mathbb E_{x_{t_2}}} [\kappa(x_{t_1},x_{t_2})] ,
\end{split}
\end{equation}
where ${\mathbb E}[\cdot]$ denotes the expectation value and $\kappa(\cdot,\cdot)$ is any positive definite kernel \cite{Principe2010}. A kernel can be viewed as a similarity measure for the data \cite{Scholkopf2002}. The Gaussian kernel which is translation-invariant, is defined as follows:

\begin{equation} \label{Gausskernel}
	G_{\sigma}(x - z) = \frac{1}{\sqrt{2\pi} \sigma}\exp\left(-\frac{\|x-z\|^2}{2\sigma^{2}}\right),
\end{equation}
where $\sigma$ is the kernel size or bandwidth. The  kernel size can be interpreted as the resolution in which the correntropy function search for similarities in the high-dimensional kernel feature space \cite{Principe2010,Santamaria2006}. The kernel size gives the user the ability to control the emphasis given to the higher-order moments with respect to second-order moments. For large values of the kernel size, the second-order moments have more relevance and the correntropy function approximates the conventional correlation. On the other hand if the kernel size is set too small, the correntropy function will not be able to discriminate between signal and noise and approximates the Dirac delta function.

The name correntropy was coined due to its close relation to Renyi's quadratic entropy, which can be estimated through Parzen windows \cite{Principe2010} as follows:

\begin{equation} \label{RQE4}
	\hat{H}_{R2}(X)=-\log \left( IP_{\sigma}(X) \right),
\end{equation}
where
\begin{equation} \label{IP}
	IP_{\sigma}(X) = \frac{1}{N^2} \sum_{i=1}^{N}{\sum_{j=1}^{N}{ G_{\sigma}(x_i - x_j)  } },
\end{equation}
and $N$ is the number of samples of the random variable $X$ and $\sigma$ is the kernel size of the Gaussian kernel function. Equation \eqref{IP} is the argument of the logarithm in Eq. \eqref{RQE4} and is called the Information Potential (IP). The mean value of the correntropy function over the lags is a biased estimator of the IP \cite{Principe2010}.

For a discrete strictly stationary random process $\{X_n\}$, the univariate correntropy function or autocorrentropy can be defined as

\[
	V[m]=  {\mathbb E}[\kappa(x_{n},x_{n-m})],
\]
which can be estimated through the sample mean
\begin{equation}\label{autocorr}
	\widehat{V}_{\sigma}[m] = \frac{1}{N-m+1} \sum_{n=m}^{N}{ G_{\sigma}(x_{n}-x_{n-m}) }.
\end{equation}
Likewise, the estimator of the univariate centered correntropy function (Eq. \ref{centeredCorr}) is

\begin{eqnarray} \nonumber \label{centeredCorrEst}
	\widehat{U}_{\sigma}[m] &=& \frac{1}{N-m+1} \sum_{n=m}^{N}{ G_{\sigma}(x_{n}-x_{n-m}) } \\
		& &- \frac{1}{N^2} \sum_{n=1}^{N} \sum_{m=1}^{N} { G_{\sigma}(x_n-x_m) },
\end{eqnarray}
where the Gaussian kernel with kernel size $\sigma$ is used, $N$ is the number of samples of $\{X_n\}$ and  $m \in [1,N]$ is the discrete time lag. In practice, the maximum lag should be chosen so that there are enough samples to estimate correntropy at each lag. Notice that the second term in Eq. \eqref{centeredCorrEst} corresponds to the IP (Eq. \ref{IP}), which is the mean of the autocorrentropy function over the lags.

The Fourier transform of the centered autocorrentropy function is called correntropy spectral density (CSD) and is defined as \cite{Principe2010,Santamaria2006,Xu2008}:
\begin{equation}\label{CSD}
	P_{\sigma}[f] = \sum_{m=-\infty}^{\infty}{ \widehat{U}_{\sigma}[m] \cdot \exp \left(-j2\pi f \frac{m}{F_s} \right) }
\end{equation}
where $\widehat{U}_{\sigma}[m]$ is the univariate centered correntropy function, and $F_s$ is the sampling frequency. The CSD can be considered as a generalized power spectral density (PSD) function, although it is a function of the kernel size and it does not measure power. As with correntropy, the kernel size controls the influence of the higher-order moments versus the second-order statistical descriptors. Particularly, for large values of the kernel size, the CSD approximates the conventional PSD.

In \cite{Xu2008} the correntropy function and the CSD were used to solve the problem of detecting the fundamental frequency in speech signals. Correntropy outperformed conventional correlation, showing better discriminatory and robustness to noise.  In \cite{Liu2007} correntropy was used to solve the blind source separation (BSS) problem, successfully separating signals coming from independent and identically distributed sources and also Gaussian sources. Correntropy outperformed methods that also make use of higher-order statistics such as Independent Component Analysis (ICA). In \cite{Gunduz2009}, correntropy was used as a discriminatory metric for the detection of nonlinearities in time series, outperforming traditional methods such as the Lyapunov exponents.

\subsection{Periodic kernel functions}
Kernels can also be viewed as covariance functions for correlated observations at different points of the input domain \cite{Scholkopf2002}. In our research we are interested in measuring similarities among samples separated by a given period. A kernel function is periodic with period $P$ if it repeats itself for input vectors separated by $P$. Periodic kernel functions are appropriate for  nonparametric estimation, modelling and regression of periodic time series \cite{Michalak2010}. Periodic kernel functions have also been proposed in the Gaussian processes literature \cite{Rasmussen2006,Mackay1998,Wang2012}.

A periodic kernel function can be obtained by applying a nonlinear mapping (or warping) $u(z)$ to the input vector $z$. In \cite{Mackay1998} a periodic kernel function was constructed by mapping a unidimensional input variable $z$ using a periodic two-dimensional warping function defined as
\begin{equation} \label{warp}
u_P \left( z\right) = \left (\cos \left(\frac{2\pi}{P} z\right) , \sin \left( \frac{2\pi}{P} z\right) \right).
\end{equation}
The periodic kernel function $G_{P;\sigma}(z-y)$ with period $P$, is obtained by applying the warping function (Eq. \ref{warp}) to the inputs of the Gaussian kernel function (Eq. \ref{Gausskernel}). The periodic kernel function is defined as,

\begin{equation}  \label{Psinkernel}
\begin{split}
G_{\sigma;P}(z-y) = G_{\sigma}(u_P(z)-u_P(y)) \\ ~= \frac{1}{\sqrt{2\pi}\sigma} \exp \left ( - \frac{\sin^2 \left( \frac{\pi}{P}  (z - y) \right)}{0.5\sigma^2}\right ),
\end{split}
\end{equation}
where the following expression is  used
\[
\left \|u_P \left( z \right) - u_P \left( y\right)   \right \|^2 = 4 \sin^2 \left( \frac{\pi  (z - y)}{P} \right).
\]
The periodic kernel function \eqref{Psinkernel} is related to the von Mises probability density function \cite{Evans2000}.

\section{Correntropy kernelized periodogram}

In this paper we propose an ITL based method for finding periodicities in unevenly sampled time series. The proposed method does not require any resampling, slotting or folding scheme, as it is computed directly from the available samples and detects periodicity using the actual magnitudes and time instants of the samples. The new metric combines the centered correntropy function and the periodic kernel function. For a discrete unidimensional random process $\{X_n\}$ with $n=1,\ldots,N$, kernel sizes $\sigma_t$ and $\sigma_m$, and a trial period $P_t$, the proposed metric is computed as
\begin{equation}   \label{HP}
\begin{split}
\widehat{V}_{P\{\sigma_t,\sigma_m\}}(P_t) =    \frac{\sigma_m}{N^2}  \sum_{i=1}^N \sum_{j=1}^N  \left( G_{\sigma_m}(\Delta x_{ij}) -IP \right) \\
\cdot ~ G_{\sigma_t;P_t}( \Delta t_{ij}) \cdot w( \Delta t_{ij}),
\end{split}
\end{equation}
where $\Delta x_{ij} = x_i - x_j$, $\Delta t_{ij} = t_i - t_j$,  $G_{\sigma_m}(\cdot)$ is the Gaussian kernel function (Eq. \ref{Gausskernel}), IP is the information potential (Eq. \ref{IP}), $G_{\sigma_t;P_t}(\cdot)$ is the periodic kernel function (Eq. \ref{Psinkernel}) and $w(\Delta t_{ij})$ is a Hamming window.

In Eq. \eqref{HP} magnitude differences are evaluated using the Gaussian kernel function, while time differences are evaluated using the periodic kernel function. The new metric performs the pointwise multiplication between the centered Gaussian kernel coefficient $\left(G_{\sigma_m}(\Delta x_{ij})-IP \right)$ and the periodic kernel coefficients $G_{\sigma_t;P_t}(\Delta t_{ij})$. Notice that Eq. \eqref{HP} is a function of the trial period $P_t$  and it has two free parameters: the magnitude kernel size $\sigma_m$ and the time kernel size $\sigma_t$. By analogy with the conventional periodogram (square of the discrete Fourier transform of the data) we call Eq. \eqref{HP} the Correntropy Kernelized Periodogram (CKP). Another extension of the conventional theory is the wavelet periodogram \cite{Abramovich2000}. A Hamming window defined as
\begin{equation} \label{hamming}
	w(\Delta t_{ij}) = 0.54 + 0.46 \cdot \cos \left(\frac{\pi \Delta t_{ij}}{T} \right),
\end{equation}
where $T$ is the total time span of the light curve, is used in Eq. \eqref{HP} to have a smoother estimation of the periodogram \cite{Jenkins1968}. 

The periodic kernel gives larger weights to the samples whose time difference are multiples of the the trial period. In other words, the periodic kernel emphasizes the magnitude differences that are separated by the trial period and its multiples. Intuitively, for a periodic time series, the magnitude differences between samples separated by the underlying period in time are expected to be minimum (most similar). If the trial period of the periodic kernel is set to the true period,  the metric is expected to reach a maximum value. By using the maximum value of Eq. \eqref{HP} over a set of trial periods, the best estimation of the underlying period is obtained.

In order to compare periodograms from different time series, the CKP has to be invariant to the data scale. In
\cite{Principe2010} a scale-invariant criterion based on Renyi's quadratic entropy is proposed. The condition for Gaussian variables is that the magnitude's kernel size should be directly proportional to the spread of the data. Herein, we use this condition as an approximate method to make the representation scale invariant. In Eq. \eqref{HP} we set $\sigma_m$ as follows,
\begin{equation} \label{sigmarule}
\sigma_m (X_n) = k \cdot \min \left\{0.7413 ~\text{iqr}(X_n), \text{std}(X_n)\right\},
\end{equation}
where $k$ is a constant, iqr is the interquartile range\footnote{If the data does not follow a normal distribution or it contains outliers, the interquartile range will provide a  better spread estimation than the standard deviation because it uses the middle 50\% of the data} and std is the standard deviation \cite{Principe2010}. The iqr is a measure of statistical dispersion, being equal to the difference between the third and first quartiles of the data. The first and third quartiles are the medians of the first half and second half of the rank-ordered data, respectively. For a normal distribution the standard deviation is equal to $0.7413 ~\text{iqr}$. The selection of constant $k$ is discussed in Section \ref{selectk}.

\begin{figure}[t]
  \centering
	\subfloat[]{\label{fig-example1a}\includegraphics[scale=0.55]{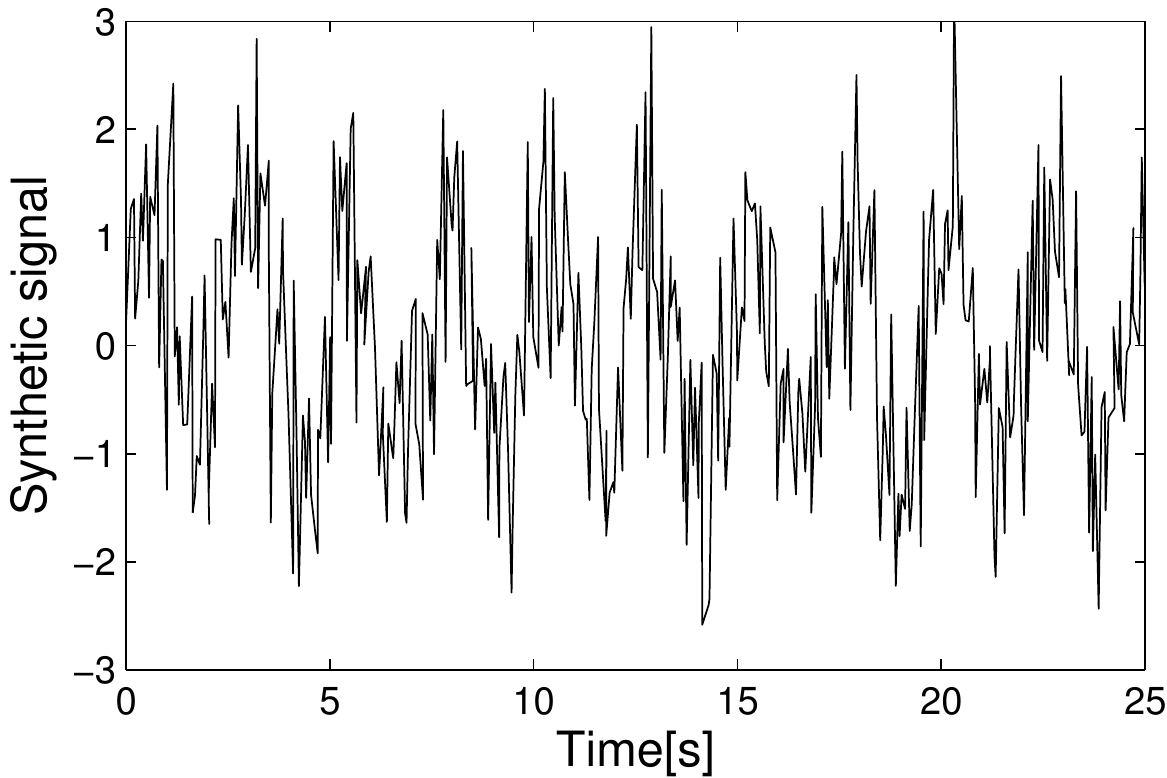} }  \qquad
	\subfloat[]{\label{fig-example1b}\includegraphics[scale=0.55]{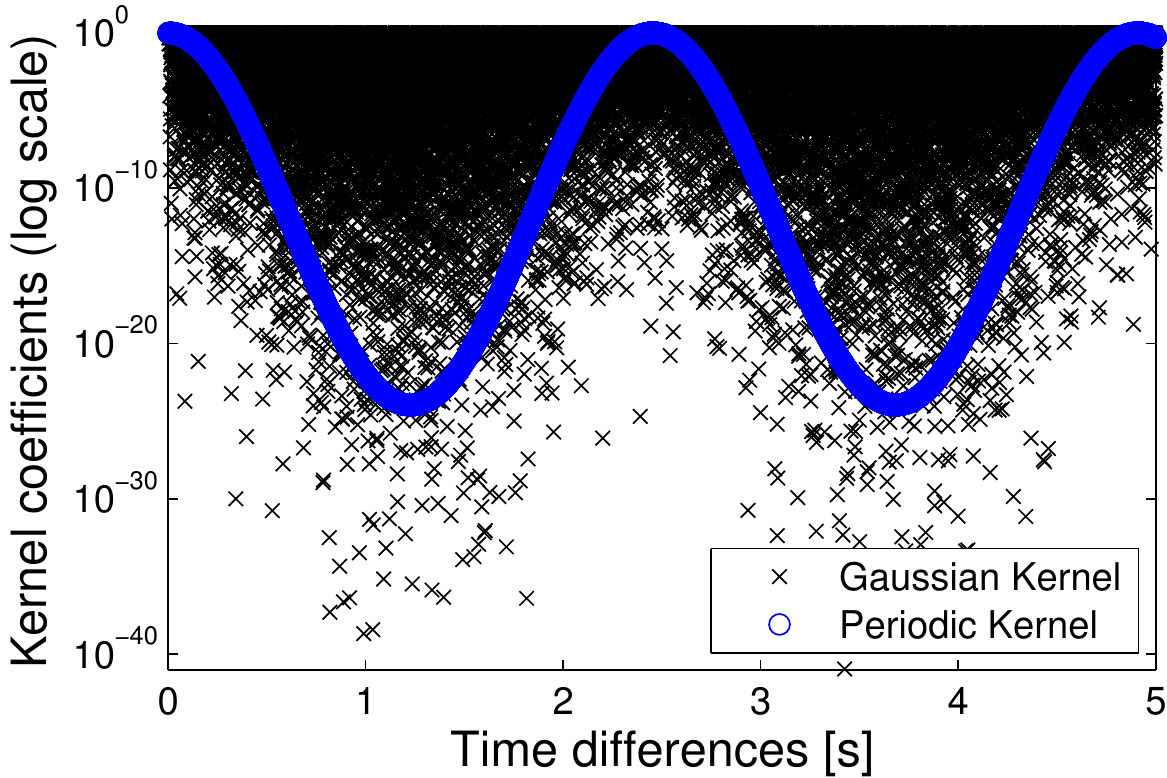} }  \qquad
	\subfloat[]{\label{fig-example1c}\includegraphics[scale=0.55]{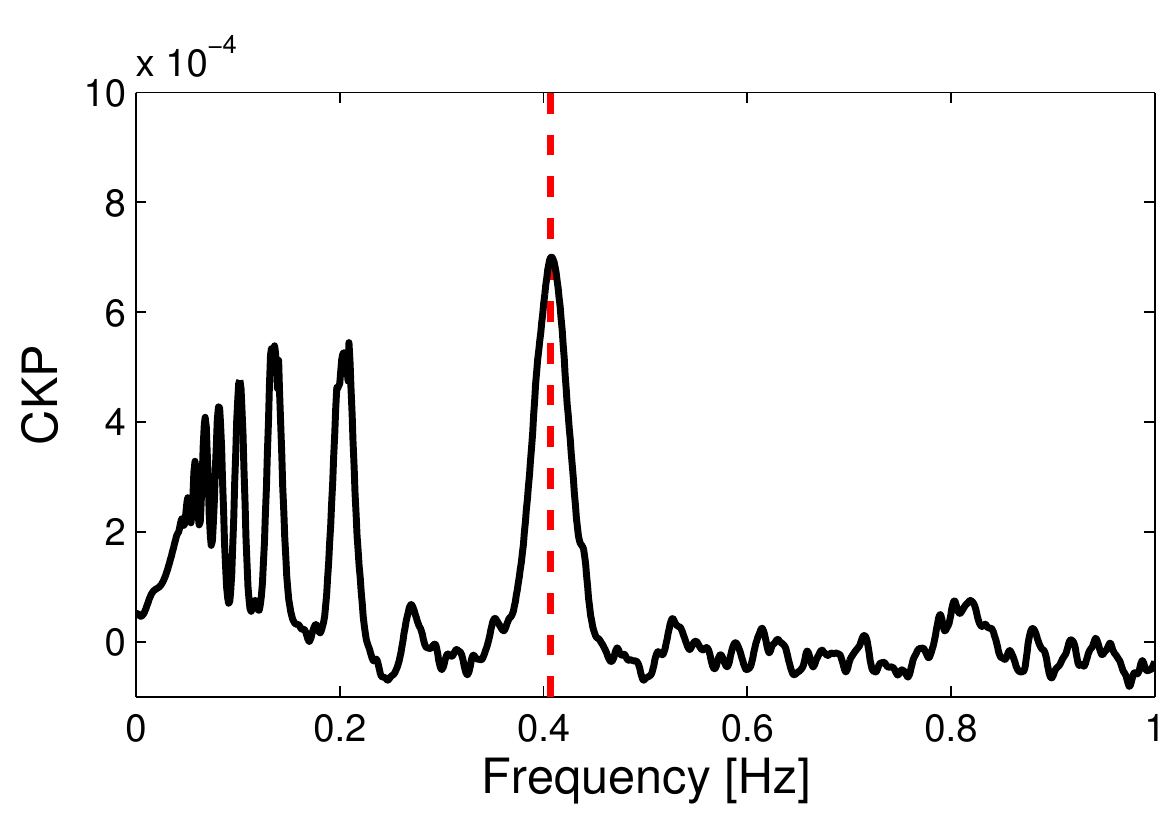} }
  \caption{ (a) Synthetic time series $sin(2\pi t/P)$ with period $P=2.456$ seconds plus Gaussian noise. The time instants have been randomly perturbed to simulate uneven sampling. (b) Kernel coefficients, $\left(G_{\sigma_m}(\Delta x_{ij}(\Delta t_{ij}))-IP \right)$ and $G_{\sigma_t;P_t}(\Delta t_{ij})$, as a function of the time differences. The CKP is the pointwise product between the centered Gaussian kernel coefficients and the periodic kernel coefficients.  (c) CKP as a function of the trial frequency, the dotted line marks the location of the true period. The global maximum of the CKP corresponds to the underlying period.} \vspace{-12pt}
\end{figure}

Fig. 1 shows an example using a synthetic time series to illustrate the effect of the proposed metric. Fig. \ref{fig-example1a} shows a synthetic time series $y_i=\sin(2\pi t_i/P) + 0.8 \cdot\epsilon_i$, with $t_i= \frac{T_{max}}{N} (i  + 0.5 \cdot \varepsilon_i$), where $\epsilon_i$ and $\varepsilon_i$ are normally distributed random variables with zero mean and unit standard deviation. The noise in time simulates uneven sampling. In this example $N=400$, $T_{max}=25$, and the underlying period is $P=2.456$ seconds. Fig. \ref{fig-example1b} shows the kernel coefficients $\left(G_{\sigma_m}(\Delta x_{ij}(\Delta t_{ij}))-IP \right)$ and $G_{\sigma_t;P_t}(\Delta t_{ij})$ as a function of the time differences collected from the time series. The magnitude kernel size is set using Eq. \eqref{sigmarule} with $k=0.3$. The time kernel size and the trial period are set to $\sigma_t=0.1$, $P_t=2.456$, respectively. Fig. \ref{fig-example1c} shows the CKP for a range of periods, the location of the underlying period in the periodogram is marked with a dotted line. The CKP reaches a global maximum at the corresponding underlying period $P=2.456$.

\subsection{A statistical test based on the CKP for periodicity discrimination} \label{statisticaltest}

For a periodic time series  with an oscillation frequency $f$,  its periodogram  will exhibit a peak at that frequency with high probability. But the inverse is not necessarily true, a peak in the periodogram does not imply that the time series is periodic. Spurious peaks may be produced by measurement errors, random fluctuations, aliasing or noise.

In this section, a statistical test for periodicity is introduced, using the global maximum of the CKP as test statistic. The null hypothesis is that there are no significant periodic components in the time series. The alternative hypothesis is that the CKP maximum corresponds to a true periodicity. The distribution of the maximum value of the CKP is obtained through Monte-Carlo simulations. Surrogate time series \cite{Schmitz1999,Schreiber1999} are used to test the null hypothesis. The surrogate generation algorithm has to be consistent with the null hypothesis. To achieve this, the block bootstrap method \cite{Buhlmann99}, which breaks periodicities preserving the noise characteristic and time correlations of the light curve, is used. The procedure used to construct an unevenly sampled surrogate using the block bootstrap method is as follows

\begin{enumerate}
\item Obtain a data block from the light curve by randomly selecting a block of length $L$ and a starting point $j\in [1,N-L]$.
\item Subtract the first time instant of the block, so that it starts at 0 days.
\item Add the value of the last time instant of the previous block to the time instants of the current block.
\item Parse the current block to the surrogate time series.
\item Repeat steps 1-4 until the surrogate time series have the same amount of samples of the original light curve.
\end{enumerate}

For a given significance level $\alpha$ and kernel sizes $\sigma_t$ and $\sigma_m$, the null hypothesis is rejected if
\[
\max_f \widehat{V}_{p\{\sigma_m,\sigma_t\}} (f) > \widehat{V}_{p\{\sigma_m,\sigma_t\}}^{\alpha},
\]
where for N light curves $\widehat{V}_{p\{\sigma_m,\sigma_t\}}^{\alpha}$ is pre-computed as follows:
\begin{enumerate}
\item Generate $M$ surrogates from each light curve using block bootstrap.
\item Save the maximum CKP ordinate value  of each surrogate.
\item Find $P_\alpha$ such that a $(1-\alpha)$\% of the ordinate values saved from the surrogates are below this threshold (one-tailed distribution).
\item Compute $\widehat{V}_{p\{\sigma_m,\sigma_t\}}^{\alpha}$ as the  mean $P_\alpha$  and its corresponding error bars as the standard deviation of $P_\alpha$ for the $N$ light curves ($N\cdot M$ surrogates).

\end{enumerate}

\section{Period Detection Method}

\subsection{Description of the MACHO database}

The MACHO project \cite{Alcock2000} was designed to search for gravitational microlensing events in the Magellanic Clouds and the galactic bulge. The project started in 1992 and concluded in 1999. More than 20 million stars were surveyed. The MACHO project has been an important source for finding variable stars. The complete light curve database is available through the MACHO project's website\footnote{http://wwwmacho.anu.edu.au/}. There are two light curves per stellar object: channels blue and red. Only the blue channel light curves are used here. Each light-curve has approximately 1000 samples and contains 3 data columns: time, magnitude and an error estimation for the magnitude.

Astronomers from the Harvard Time Series Center (TSC) have a catalog of variable stars from the MACHO survey. The underlying periods of the periodic variable stars were estimated using epoch folding, AoV, and visual inspection. In this paper, we consider the TSC periods to be the gold standard.

A subset of 1500 periodic light curves (500 Cepheids, 500 RR Lyrae and 500 eclipsing binaries) and 3500 non-periodic light curves was drawn from the MACHO survey. The subset was divided into a training set for parameter adjustment and a testing set. The training set consisted of 2500 light curves (750 periodic and 1750 non periodic) randomly selected from the available classes. The remaining 2500 light curves were used for testing  purposes.

There is a natural imbalance between periodic and aperiodic classes of stars. Only 3\% of the surveyed stars are expected to be variables and $\sim$1\%  to be periodic. Due to this, when detecting periodic behaviour, we have to achieve a false positive rate less than 0.1\%.

\subsection{Description of the procedure for periodicity detection} \label{pipeline}

In what follows, the steps of the periodicity detection algorithm, for a given time kernel size $\sigma_t$ and magnitude kernel size $\sigma_m$, are described.

\begin{enumerate}
\item \textbf{Cleaning:} The light curve's blue channel is imported. The mean $\bar{e}$ and the standard deviation $\sigma_e$ of the photometric error are computed. Samples that do not comply with $e_i < \bar{e} + 3\cdot \sigma_e$, where $e_i$ is the photometric error of sample $i$, are discarded.
\item \textbf{Computing the periodogram} The CKP (Eq. \ref{HP}) is computed on 20000 logarithmically spaced periods between 0.4 days and 300 days. The periods associated to the ten highest local maxima  at the periodogram are saved as trial periods for the next step.
\item \textbf{Fine-tuning of trial periods:} The CKP is used to fine tune the ten trial periods. Each trial period is fine-tuned around a 0.5\% of its value ($[1.0025\cdot f_{trial},0.9975\cdot f_{trial}]$), using a step size of $df=\frac{0.01}{T}$ in frequency, where T is the total time span of the light curve.
\item \textbf{Selection of the best trial period:} The trial periods are sorted in descending order following its CKP ordinate value. The best trial period $P_{best}$ is selected as the one with the highest value of $\widehat{V}_P$, that is not a multiple of a spurious period, as described below.
\item Finally, if the best period comply with $\widehat{V}_{P\{\sigma_m,\sigma_t\}}(P_{best}) > \theta$ then the light curve is labeled as periodic, where $\theta$ is the periodogram threshold for periodicity. The confidence associated to $\theta$ is obtained using the procedure described in Section \ref{statisticaltest}.
\end{enumerate}
Table \ref{tab-case1} gives a summary of the parameters of the proposed method. The kernel sizes and the periodicity threshold do not appear in Table \ref{tab-case1}, because they need to be calibrated using a procedure described in the following sections.

To obtain the spurious periods the following spectral window function is used
\begin{equation} \label{SW}
W(f) = \frac{1}{N} \left | \sum_{i=1}^N \exp \left(j 2 \pi f t_i \right)\right|^2,
\end{equation}
where $t_i$ with $i=1,\ldots,N$ are the time instants of the light curve. Eq. \eqref{SW} is the periodogram of the sampling pattern of the light curve. The frequencies associated to the peaks of the spectral window are related to spurious periodicities caused by the sampling. In most cases, the spurious periods obtained from the spectral window are multiples of the sidereal day ($0.99727$ days) and yearly periodicities ($365.25$ days). The moon phase period (29.53 days) is also added to the list of spurious periods. The moon phase has no relation to the sampling but it is intrinsic to the data. The daily sampling produces alias peaks of the true periods (P) in the CKP at $(1/P+k)$ 1/days, where $k\in \mathbb{Z}$. The CKP values of the aliases are very low and therefore they do not need to be filtered out.

\addtocounter{footnote}{1}
\footnotetext[\value{footnote}]{Logarithmically spaced.}
\addtocounter{footnote}{1}
\footnotetext[\value{footnote}]{Linearly spaced, T is the total time span of the light curve.}
\addtocounter{footnote}{-1}

\begin{table}[t]
	
	\caption{ \label{tab-case1} Summary of parameters of the period detection pipeline }
	\begin{center}
		\begin{tabular}{  l  | c  }
		Description & Value \\ \hline
		Minimum period  &  0.4 days \\
		Maximum period & 300 days \\
		Periodogram resolution & 20.000 periods$^{\decimal{footnote}}$\addtocounter{footnote}{1}  \\
		Number of fine-tuned trial periods & 10 \\	
		Fine-tune resolution & $\frac{0.01}{T}$ periods$^{\decimal{footnote}}$\\

		\end{tabular} \vspace{-15pt}
	\end{center}
\end{table}

\subsection{Computational time}

The algorithm\footnote{The C-language implementation is available by request to the authors.} was programmed in C/CUDA and implemented in a Graphical Processing Unit (GPU) NVIDIA Tesla C2050 with 448 cuda cores. The computational time required to process one light curve (1000 samples) is 1.812 seconds. The total time to process the 5000 light curves is 2.5 hours.

\subsection{Performance criteria}

In what follows we define performance measures for the problems of period detection and period estimation. The former consists of discriminating between periodic and non periodic light curves. The latter consists in estimating the true periods of periodic light curves. For the period estimation problem the classification is done by using the TSC periods as golden standard. An estimated period $P_{est}$ is classified as either a Hit, a Multiple or a Miss with respect to the reference period $P_{ref}$ according to the following criteria:
\begin{itemize}
\parskip 0pt
\item Hit if $\left|P_{ref} - P_{est} \right| < \varepsilon \cdot P_{ref}$
\item Multiple if $P_{est} > P_{ref}$ and
\[
	\left|\frac{P_{est}}{P_{ref}} - \left\lfloor \frac{P_{est}}{P_{ref}} \right\rfloor \right| < \varepsilon,
\]
or if $P_{est} < P_{ref}$ and
\[
 \left|\frac{P_{ref}}{P_{est}} - \left\lfloor \frac{P_{ref}}{P_{est}} \right\rfloor \right| < \varepsilon,
\]
where $\lfloor x \rfloor$ is the largest integer less than or equal $x$.
\item Miss if it does not belong to any of the other categories.
\end{itemize}

The tolerance value $\varepsilon$ controls the accepted relative error between the estimated period and the reference period. A value of $\varepsilon=0.005$, \emph{i.e.} a relative error of 0.5\% will be considered, small enough to obtain a clean folded curve from the estimated period.

The problem of period detection in light curves can be treated as a binary classification problem. The classes are periodic light curves $\{+1\}$ and non-periodic light curves $\{-1\}$. Confusion matrix and Receiver Operating Characteristic (ROC) curves are used to evaluate the period detection method. An ROC curve is a plot of the true positive rate (TPR) as a function of the false positive rate (FPR). Different points in the ROC curve are obtained by changing the threshold value at the output of the classifier. In this case,
\begin{itemize}
\item true positive (TP) is a periodic light curve classified as periodic
\item false positive (FP) is a non-periodic light curve classified as periodic
\item true negative (TN) is a non-periodic light curve classified as non-periodic
\item false negative (FN) is a periodic light curve classified as non-periodic
\end{itemize}

The TPR represents the proportion of periodic light curves that are correctly identified as such. The FPR represents the proportion of non-periodic light curves that are incorrectly classified as periodic.

\section{Experiments}

\subsection{Parameter calibration} \label{selectk}

\begin{figure}[t]
  \centering
\hspace*{-15pt} \includegraphics[scale=0.52]{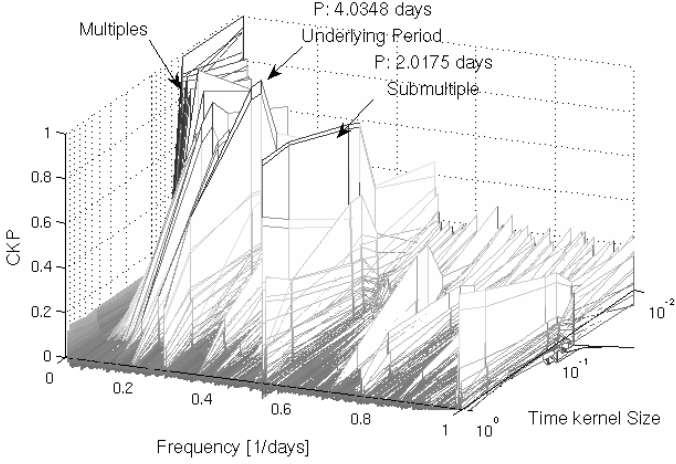}
  \caption{ \label{waterfall} CKP versus frequency and time kernel size $\sigma_t$ for light curve 1.3449.27 from the MACHO catalog. The underlying period of $4.0348$ days is found as the maximum of the CKP with $\sigma_t \in[0.075,0.125]$.  } \vspace{-10pt}
\end{figure}

The CKP is a function of the kernel sizes $\sigma_m$ and $\sigma_t$. These parameters are adjusted using the 2500 light curves in the training set. The value of $\sigma_m$ is set using Eq. \eqref{sigmarule}, so we need to choose the value of the constant $k$. Fig. \ref{waterfall} shows the CKP as function of the frequency and $\sigma_t$ for light curve 1.3449.27 from the MACHO catalog. In this example the underlying period is picked as the global maximum of the CKP for a time kernel size $\sigma_t \in[0.075,0.125]$. After extensive experiments we identified that this particular range of kernel sizes values gives the best results for the MACHO light curves. There is no clear rule for choosing the time kernel size, although intuitively, it should depend on the sampling pattern.

The period estimation results for different combinations of $k$ and $\sigma_t$ on the 750 periodic light curves of the training set are shown in Table \ref{tab:train}. The best performance is obtained with $k=1$ and $\sigma_t=0.1$.

\begin{figure}[t]
  \centering
\hspace*{-15pt} \includegraphics[scale=0.5]{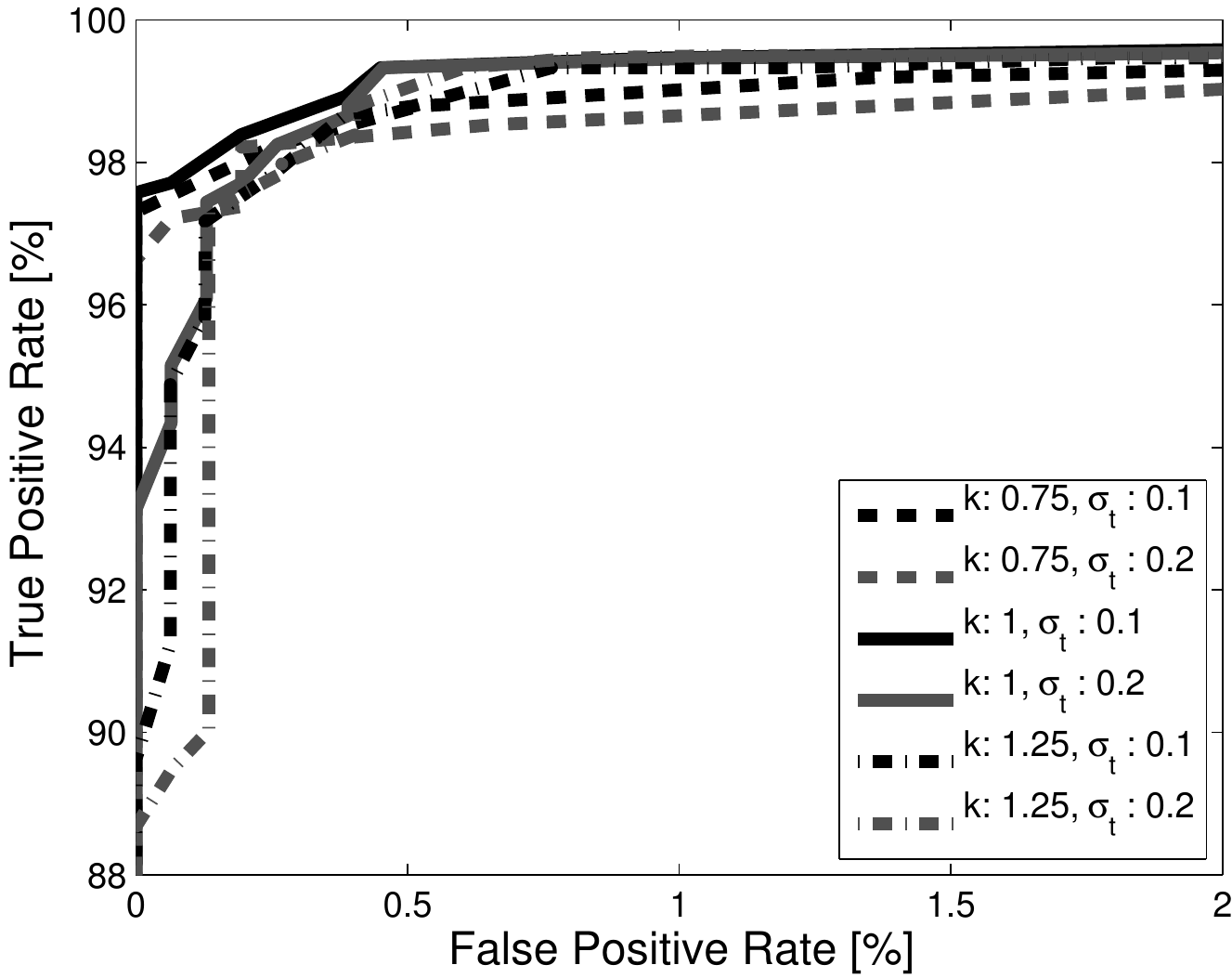}
  \caption{ \label{ROCtrain} Receiver operating characteristic curves for the period detection problem using different combinations of kernel sizes in the training database.} \vspace{-10pt}
\end{figure}

Fig. \ref{ROCtrain} compares the ROC curves obtained using different combination of both kernel sizes for the period detection problem. As mentioned before, due to the imbalance between periodic and non-periodic light curve classes, false positive rates below 0.1\% are desired. Looking at the ROC curves it is clear that the best kernel size combination, in the area below 1\% FPR, is $k=1$ and $\sigma_t=0.1$. These values are fixed for the following experiments.

\begin{table}[t]
	\begin{center}
	\caption{Period estimation performance of the CKP using different combinations of kernel sizes for the  training database}  	
	\begin{tabular}{l l c c c} \hline
		$k$ & $\sigma_t$ & Hits[\%]& Multiples[\%]& Misses[\%] \\ \hline
		$0.75$	&	$0.05$	&	$75.47$ &	$23.47$ &	$1.07$ \\
		$0.75$	&	$0.1$	&	$86.13$ &	$13.60$ &	$0.27$ \\
		$0.75$	&	$0.15$	&	$85.33$ &	$13.87$ &	$0.80$ \\
		$1$	&	$0.05$	&	$82.27$ &	$16.93$ &	$0.80$ \\
		\textbf{1}	&	\textbf{0.1}	&	\textbf{88.67} &	\textbf{11.20} &	\textbf{0.13} \\
		$1$	&	$0.15$	&	$86.93$ &	$12.67$ &	$0.40$ \\
		$1.25$	&	$0.05$	&	$76.40$ &	$22.13$ &	$1.47$ \\
		$1.25$	&	$0.1$	&	$85.73$ &	$14.13$ &	$0.13$ \\
		$1.25$	&	$0.15$	&	$85.47$ &	$14.00$ &	$0.53$ \\ \hline
	\end{tabular}
	\label{tab:train}
	\end{center} \vspace{-10pt}
\end{table}

\subsection{Statistical significance}

\begin{figure}[t]

  \centering
	\hspace*{-15pt} \includegraphics[scale=0.55]{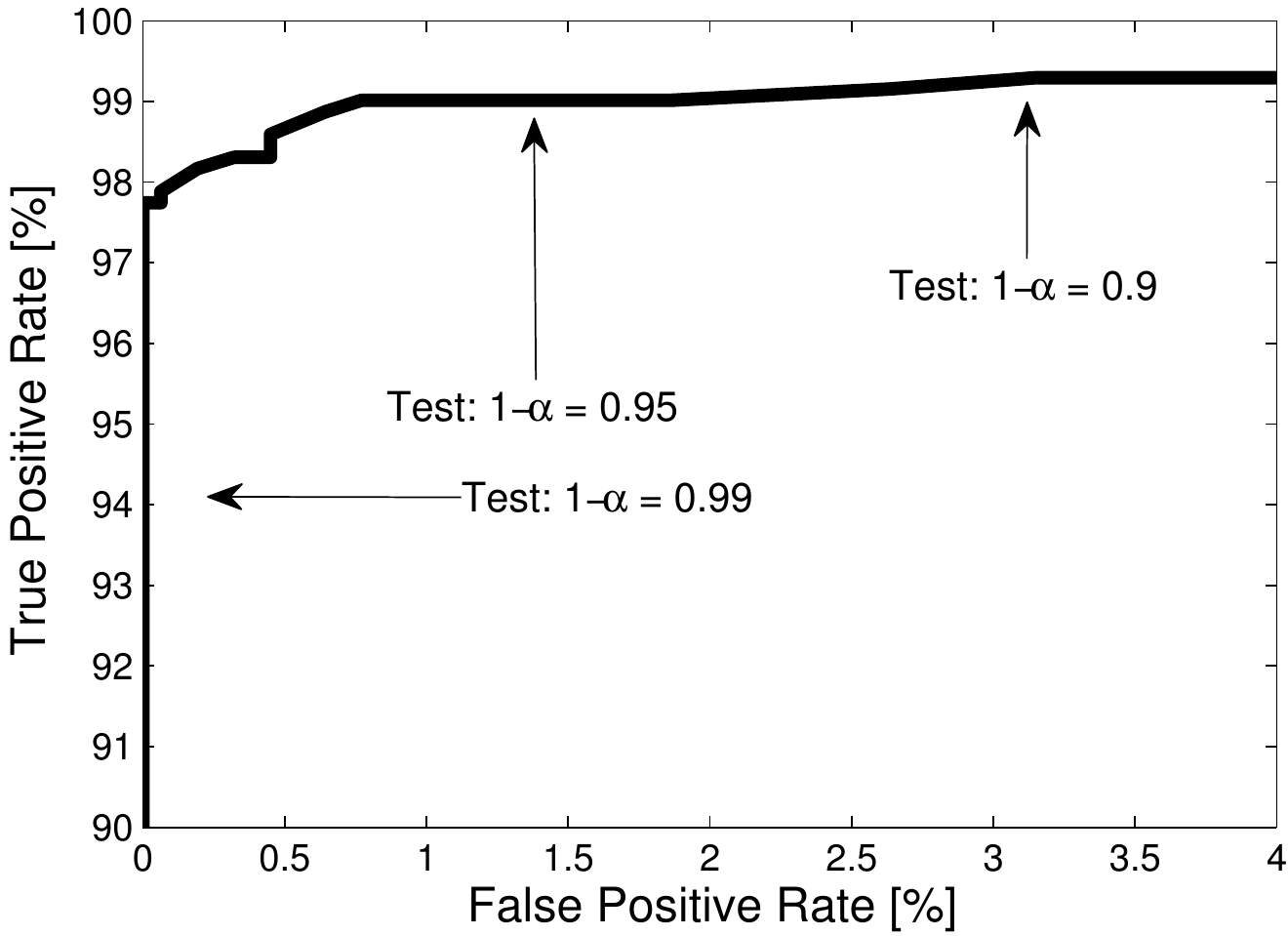} \\

  \caption{ ROC curve of the CKP with  $\sigma_t=0.1$ and $k=1$ in the test subset. The significance thresholds of the CKP are shown in the ROC curve. }
\label{fig-rocsig}
 \vspace{-10pt}
\end{figure}

\begin{figure}[t]

  \centering
\hspace*{-15pt} \includegraphics[scale=0.48]{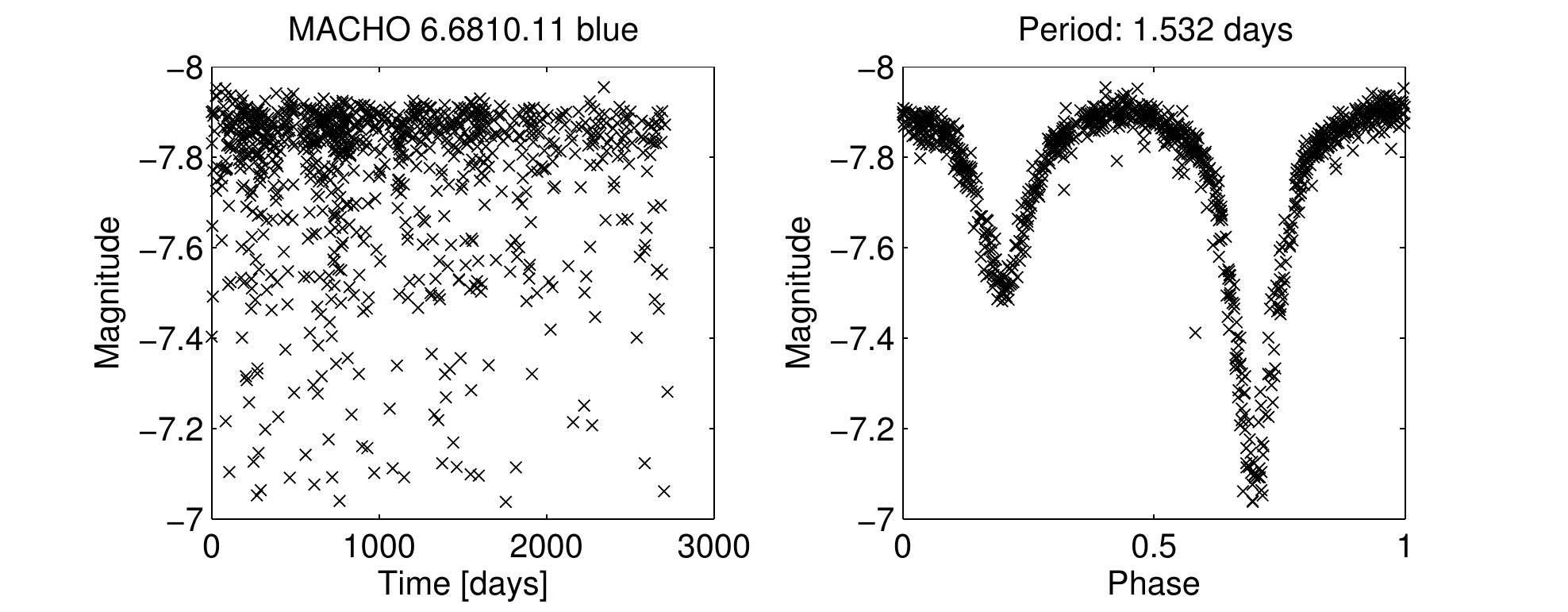} \\
\hspace*{-15pt} \includegraphics[scale=0.48]{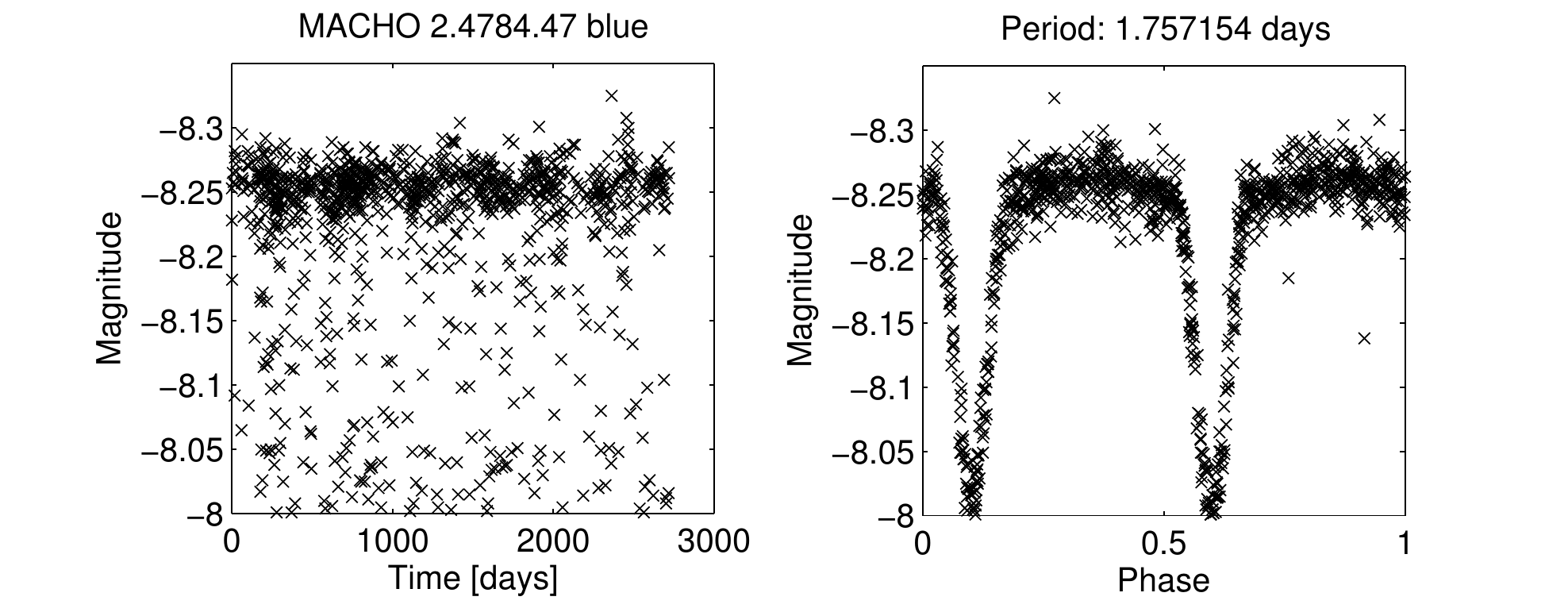} \\
\hspace*{-15pt} \includegraphics[scale=0.48]{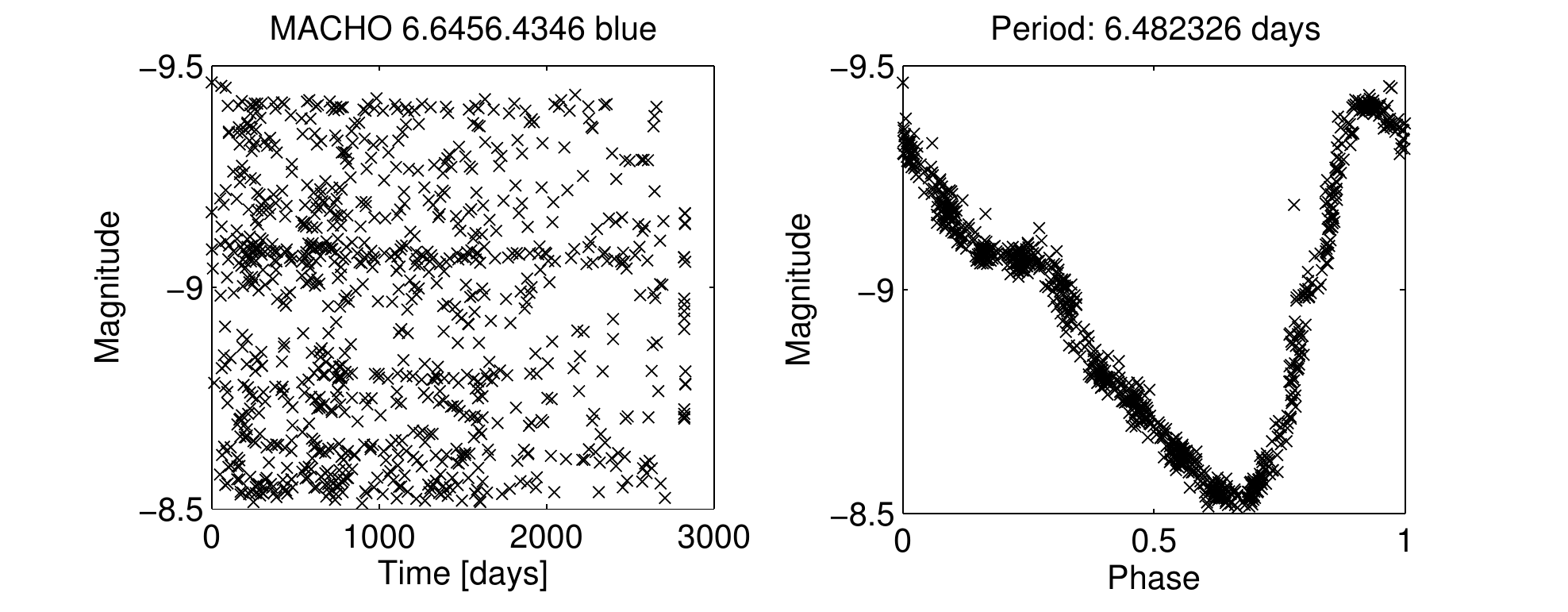}
  \caption{ Examples of periodic light curves detected by the CKP method with a level of confidence greater than 99\%. The original light curves are shown on the left column. The right column shows the same light curves folded with the estimated period. }
\label{fig-sig1}  \vspace{-18pt}
\end{figure}

\begin{figure}[t]

  \centering
\hspace*{-10pt} \includegraphics[scale=0.48]{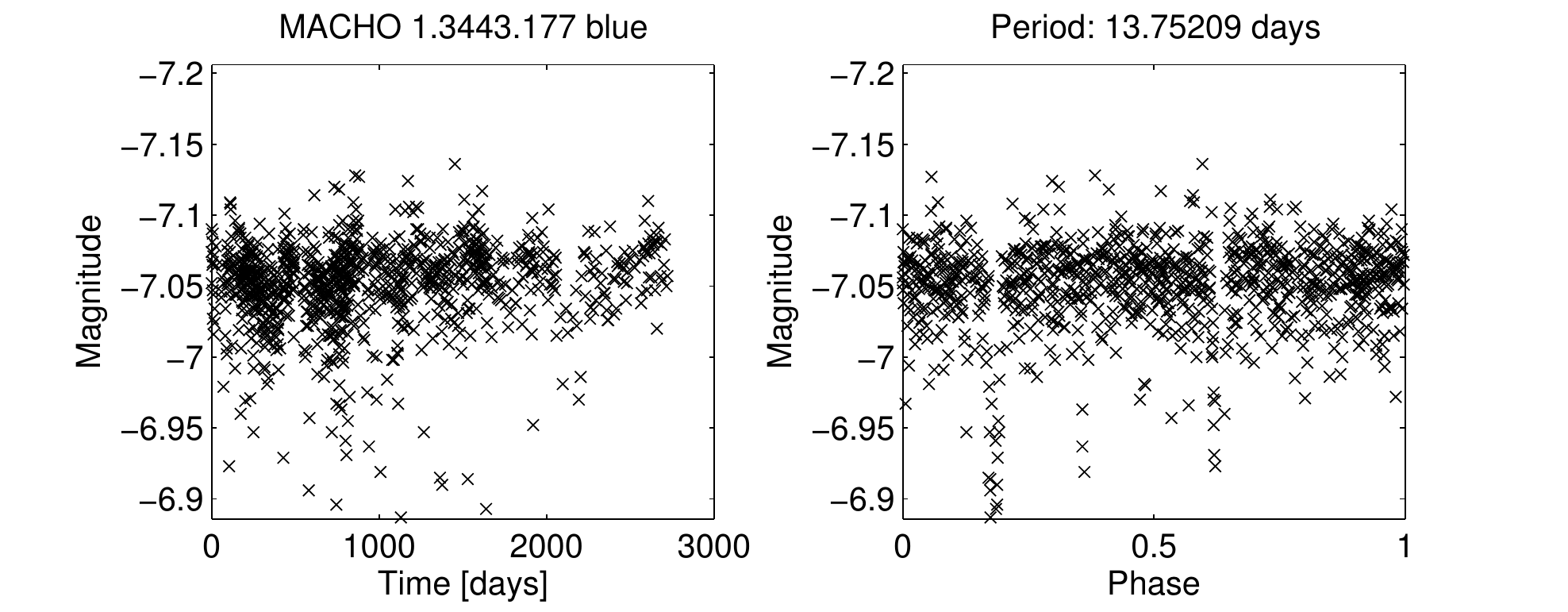} \\
\hspace*{-10pt} \includegraphics[scale=0.48]{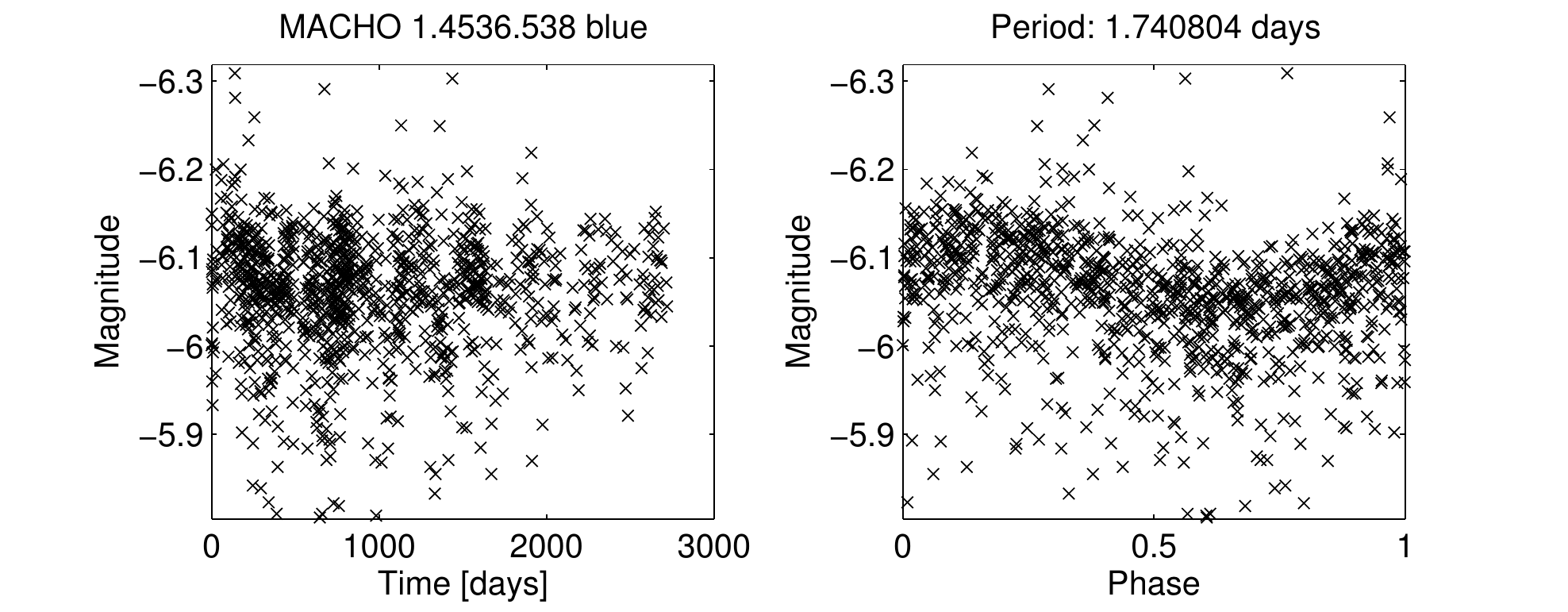} \\
\hspace*{-10pt} \includegraphics[scale=0.48]{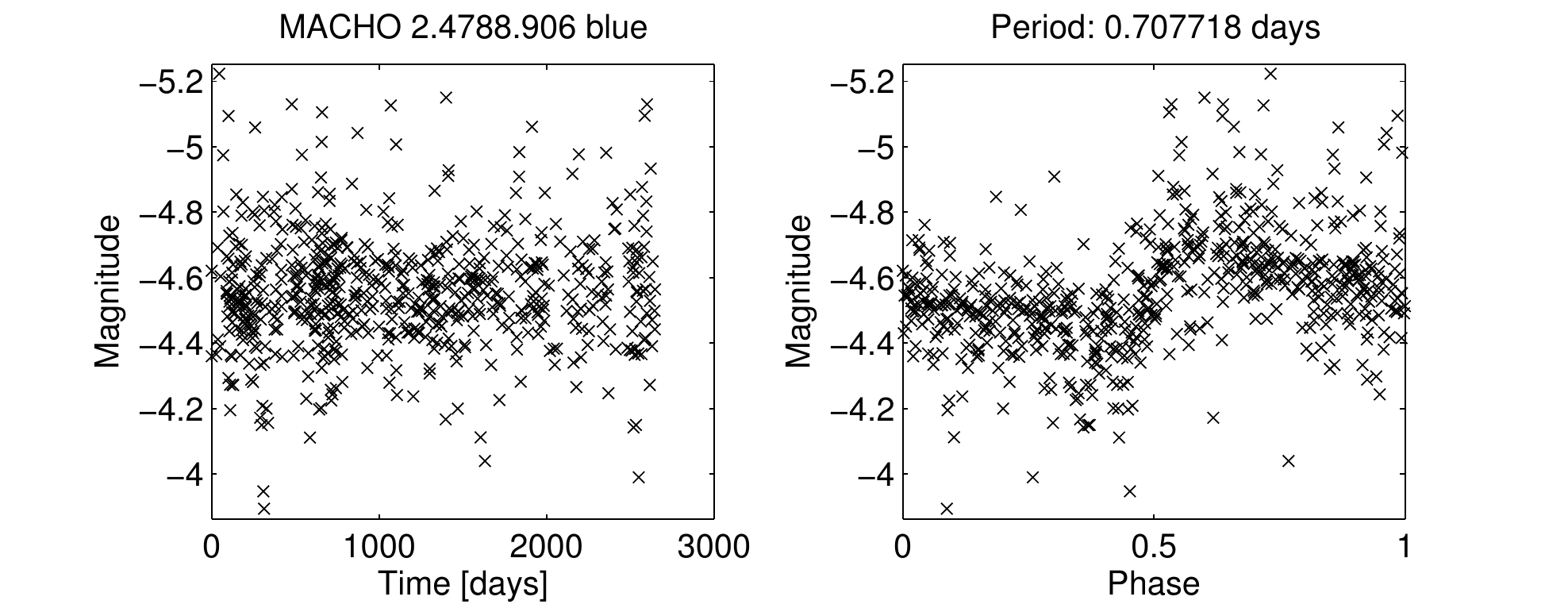}
  \caption{ Examples of periodic light curves detected by the CKP method with a level of confidence between 90\% and 95\%. The original light curves are shown on the left column. The right column shows the same light curves folded with the estimated period.  }
\label{fig-sig2}  \vspace{-5pt}
\end{figure}

\begin{figure}[t]

  \centering
\hspace*{-15pt} \includegraphics[scale=0.5]{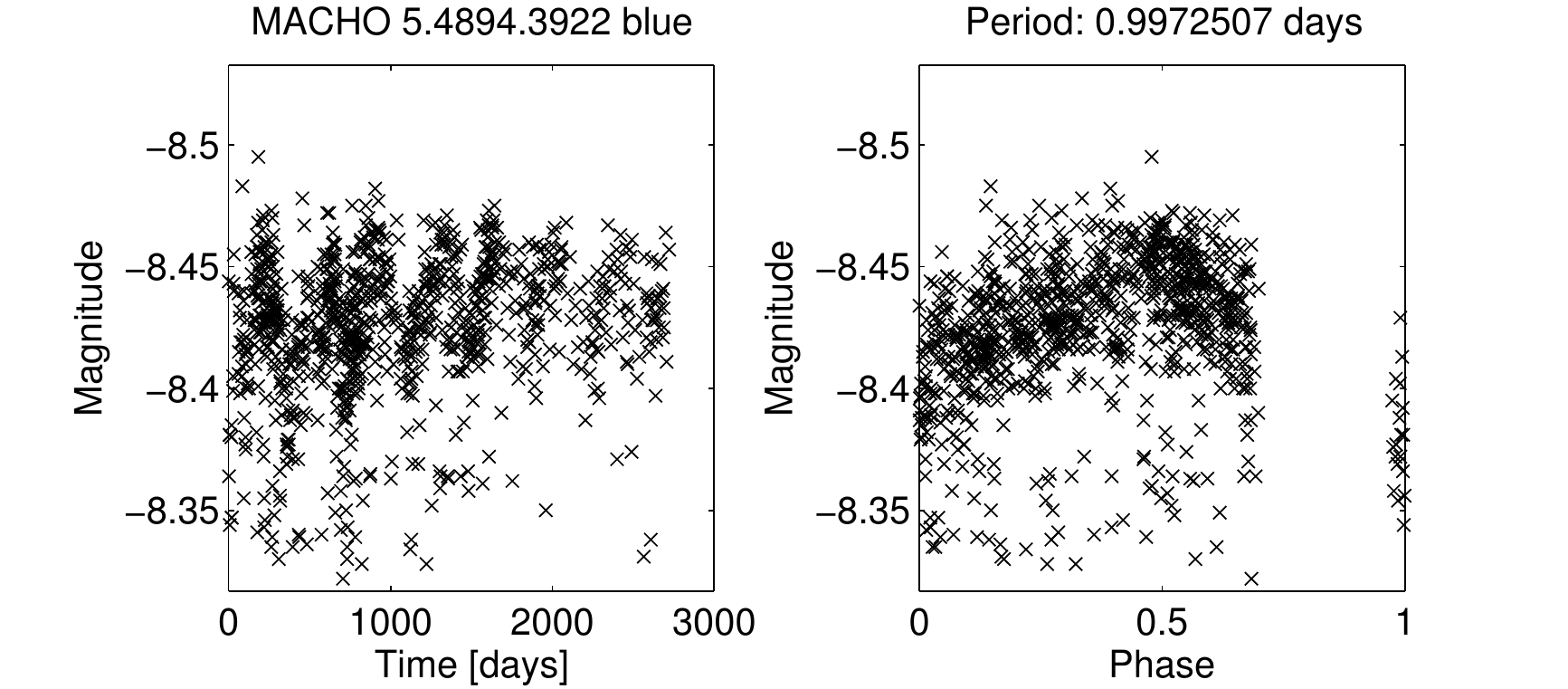} \\
\hspace*{-15pt} \includegraphics[scale=0.5]{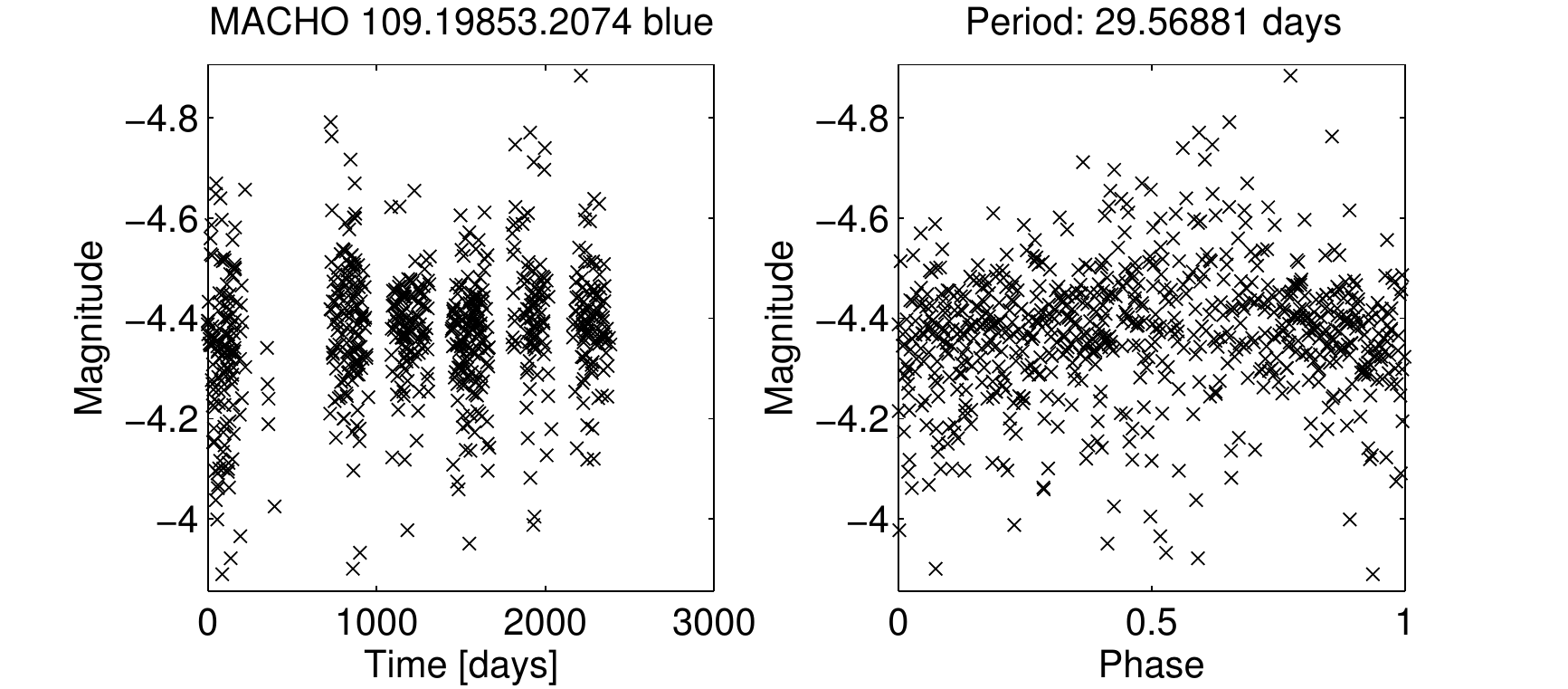}
  \caption{ Examples of spurious periods found among the top peaks of the CKP. The original light curves are shown on the left column. The right column shows the folded light curves with the spurious period. These spurious periods are related to periodic daily changes in the atmosphere and periodic monthly changes in the brightness of the moon. These periods are discarded as spurious by our method. }  \vspace{-15pt}
\label{fig-sig3}
\end{figure}

Using the procedure described in Section \ref{statisticaltest} statistical significance thresholds for the CKP were computed. Table \ref{tab:sig} shows the significance thresholds and their corresponding CKP ordinate values for the best combination of kernel sizes ($\sigma_t=0.1$ and $k=1$). The thresholds were computed using the light curve training set ($N=2500$) and five hundred surrogates per light curve ($M=500$). Fig. \ref{fig-rocsig} shows the location of these thresholds in the ROC curve of the testing dataset. FPR rates below 1\% are associated with confidence levels between 95\% and 99\%.
Fig. \ref{fig-sig1} shows three light curves in which the CKP ordinate value associated with the fundamental period has a confidence level higher than 99\%. In the folded light curves (Fig. \ref{fig-sig1} right column) the periodic nature of the light curve can be clearly observed. In period detection schemes based on visual inspection these light curves would be undoubtedly labeled as periodic. Fig. \ref{fig-sig2} shows three light curves in which the CKP ordinate values associated with the fundamental period have a statistical confidence between  90\% and 95\%. These light curves are indeed periodic although compared to the previous three (shown in Fig. \ref{fig-sig1}), their periodicity is less clear as their signal to noise ratio is smaller.
By associating a statistical level of confidence to the detected periods, we have obtained a way to set a period detection threshold and also a better interpretation of period quality. The level of confidence on the detected period could be used in later (post-processing) stages of the period detection pipeline. For example, periods with lower confidence levels may be selected for additional analysis stages in which finer resolution or different parameter combinations may be used. Fig. \ref{fig-sig3} shows examples of periods associated to the sidereal day (0.99727 days) and moon phase (29.53 days). These periods are discarded as spurious as described in Section \ref{pipeline}.


\begin{table}[t]
	\begin{center}
	\caption{Statistical significance thresholds of the CKP with $\sigma_t=0.1$ and $k=1$. }  	
	\begin{tabular}{c c c} \hline
		$1-\alpha$ & $\widehat{V}_P^{\alpha}$\\ \hline
		$0.99$	&	$3.59e-4$ \\
		$0.95$	&	$3.12e-4$ \\
		$0.90$	&	$2.80e-4$ \\

	\end{tabular}
	\label{tab:sig}
	\end{center} \vspace{-10pt}
\end{table}

\subsection{Influence of the periodic kernel}

\begin{figure}[t]
	\centering
	\includegraphics[scale=0.58]{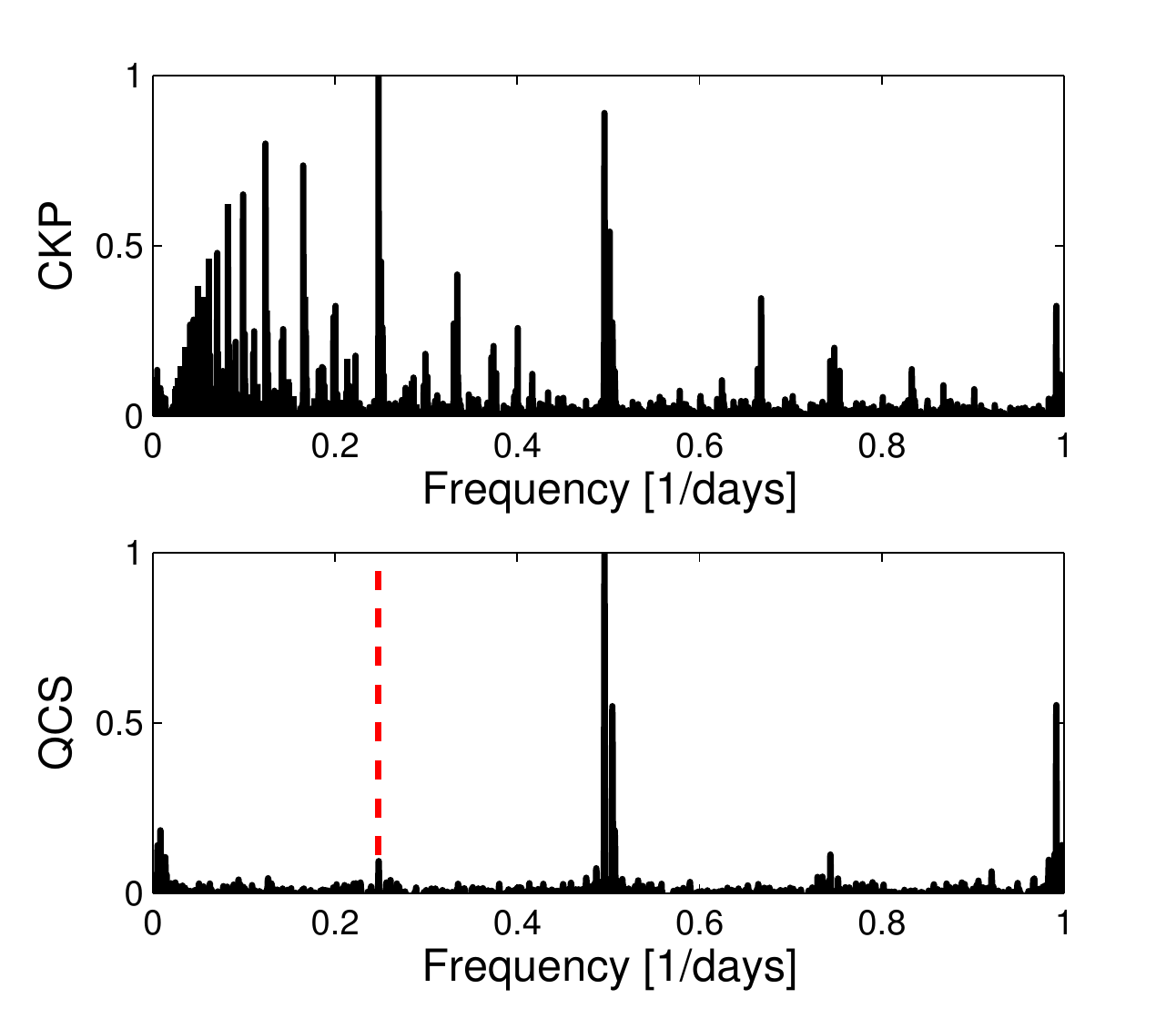}
	\caption{Comparison between the periodograms obtained using the proposed CKP metric and the QCS. The dotted line marks the true period. The underlying period corresponds to the global maximum of the CKP. On the contrary, the QCS value of the true period is very small.} \label{fig-exampleP2} \vspace{-10pt}
\end{figure}

In this experiment we demonstrate the advantages of the kernelized periodogram with respect to the conventional spectrum estimator. The quadratic correntropy spectrum (QCS) is defined as follows:
\begin{equation} \label{QCS}
\begin{split}
\widehat{P}_{\sigma_m} (f) = \sum_{i=1}^N \sum_{j=1}^N \left( G_{\sigma_m} (\Delta x_{ij}) -IP \right) \\
\cdot  \exp ( -j 2 \pi f \Delta t_{ij} ) \cdot w (\Delta t_{ij}),
\end{split}
\end{equation}
where $\Delta x_{ij}=x_i-x_j$, $\Delta t_{ij}=t_i-t_j$, and IP is the information potential estimator (Eq. \ref{IP}). Notice that Eq. \eqref{QCS} differs from  the CSD definition (Eq. \ref{CSD}) in two aspects. First the CSD definition uses integer lags, while the QCS uses directly the difference between samples. This change is useful to deal with unevenly sampled light curves. Secondly, we include in Eq. \eqref{QCS} a Hamming window (Eq. \ref{hamming}) for enhancing the spectrum estimation. The constructed quadratic correntropy spectrum (QCS) is similar to the CKP (Eq. \ref{HP}) except that the periodic kernel has been replaced by $\exp(-j 2 \pi f \Delta t)$ to compute the Discrete Fourier Transform.
First, we use an example to illustrate the differences between periodograms using a single light curve, from object 1.3449.27 of the MACHO survey. This light curve corresponds to an eclipsing binary with fundamental period $4.03488$ days.  Fig. \ref{fig-exampleP2} shows the CKP and QCS estimators for light curve 1.3449.27. The CKP was computed using $k=1$ and $\sigma_t=0.1$. The same kernel size was used for the QCS. The dotted line marks the location of the underlying period. The global maximum of the CKP is associated with $4.0349$ days, which corresponds to the underlying period of the light curve. Other local maxima appear at multiples and sub-multiples of the underlying period. On the other hand, the global maximum of the QCS is associated to $2.0175$ days, the closest integer sub-multiple of the underlying period. The QCS value at the frequency corresponding to the true period is very small, in fact smaller than many other spurious peaks. The procedure described in Section \ref{pipeline} is used to evaluate the performance of QCS estimator. Table \ref{tab:spectra} shows the results obtained by the QCS and CKP estimators on the testing subset. The CKP obtains 12\% more hits and 70\% less misses than the QCS estimator, which shows clearly the advantages of the kernelized periodogram.


\begin{table}[t]
	\begin{center}
	\caption{Period estimation performance of the CKP versus the QCS for the testing database}  	
	\begin{tabular}{l c c c}
	\hline

	\textbf{Method} & Hits[\%]& Multiples[\%]& Misses[\%] \\ \hline
	\textbf{CKP} &	\textbf{88.00} & \textbf{11.60} & \textbf{0.4} \\
	QCS  & $77.87$ & $20.80$ & $1.33$\\

	\hline
	\end{tabular}
	\label{tab:spectra}
	\end{center} \vspace{-15pt}
\end{table}

\subsection{Comparison with other methods}

\begin{figure}[t]
  \centering
 \hspace*{-15pt} \includegraphics[scale=0.58]{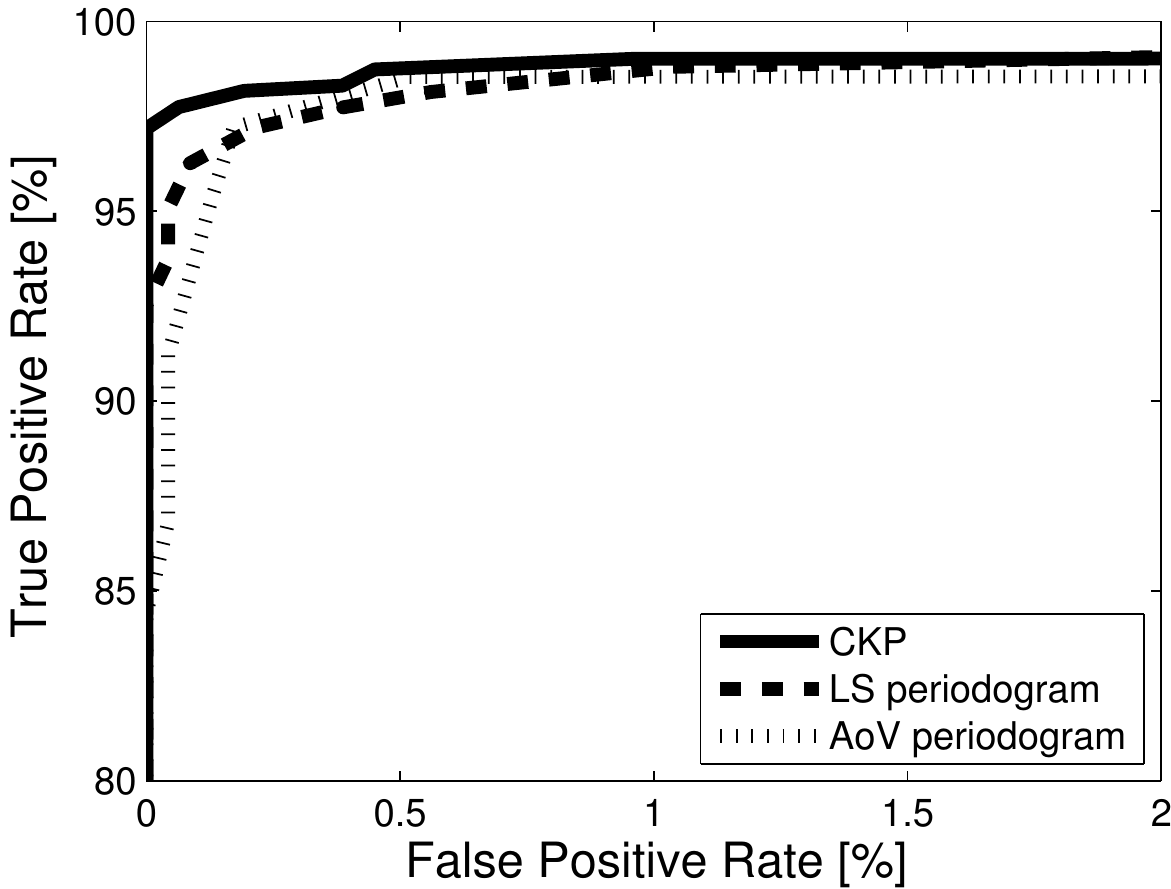}
  \caption{ \label{ROCtest} Receiver operating characteristic curves of the CKP ($k=1$, $\sigma_t=0.1$), the LS periodogram and the AoV periodogram.} \vspace{-10pt}
\end{figure}

The performance of the CKP method was compared with the slotted correntropy and other widely used techniques in astronomy. The software VarTools \cite{Hartman2008,Devor2005} was used to perform a Lomb-Scargle periodogram and Analysis of Variance analysis. The regularized Lafler-Kinman string length (SLLK) statistic and the slotted autocorrelation  were also considered.
For Vartools LS, the period associated with the highest peak of the LS periodogram, that is not a multiple of the known spurious periods (sidereal day, moon phase), gives the estimated period. A periodogram resolution of $0.1/T$ and a fine tune resolution of $0.01/T$, where T is the total time span of the light curve, were used. For Vartools AoV and SLLK, the corresponding statistics are minimized in an array of periods ranging from 0.4 to 300 days  with a step size of 1e-4. For AoV the default value of 8 bins is used. For AoV and SLLK, the period that minimizes the corresponding metrics, that is not a multiple of the known spurious periods, is selected as the estimated period.
For the slotted autocorrelation/autocorrentropy the highest peak of the PSD/CSD estimator function, that is not associated to the known spurious periods, delivers the estimated period. For the slotted autocorrentropy/autocorrelation a slot size $\Delta \tau =0.25$ was considered \cite{Huijse2011}. For the CKP method the best combination of kernel sizes ($\sigma_t=0.1$ and $k=1$) obtained with the training dataset is used.
The influence of the higher-order statistical moments is assessed by comparing the CKP with a linear version of the proposed metric. In this linear version, the Gaussian kernel used to compare magnitude values is replaced by a linear kernel. The periodic kernel remains unchanged. The procedure to detect a period is the same as explained in Section \ref{pipeline}, with an additional pre-processing step where the data vector is zero-mean normalized.
The results for period estimation on the testing subset are shown in Table \ref{tab:Final}. The CKP method obtained the highest hit rate (88\%), followed by the linear version of the CKP (80\%), the slotted correntropy (78\%) and the AoV periodogram (75\%). The CKP obtained 8.6\% more hits and 57\% less misses than its linear version. This is because the Gaussian kernel incorporates all the even-order moments of the process and gives robustness to outlier data samples which are common in astronomical time series.
The CKP obtained 10.5\% more hits and 72.7\% less misses than the slotted correntropy. This is because in the slotted correntropy, kernel coefficients are averaged on time slots, therefore the actual time differences between samples are not used.
Out of the cases where the correct period is found by the CKP but not by the AoV periodogram, an 84\% corresponds to eclipsing binaries. This is expected as conventional methods perform well on pulsating variables \cite{Huijse2011}. Eclipsing binaries light curves are typically more difficult to analyze as their variations are non-sinusoidal and due to their morphology/shape characteristics most methods tend to return harmonics or sub-harmonics rather than the true period.
The proposed method obtained the lowest miss rate (0.4\%). In all these missed cases\footnote{Light curves 1.4539.37, 3.6605.124 and 6.5726.1276, with periods 2.9955 (three times sidereal day), 3.9813 (four times the sidereal day) and 0.99676, respectively.}, the true period was correctly found by the proposed metric but they were filtered out for being too near to the sidereal day. More accurate ways of filtering out spurious periods, using the data samples instead of a straightforward comparison of the detected periods, could be implemented to recover such missed cases.
Fig. \ref{ROCtest} shows ROC curves for the task of periodic versus non-periodic discrimination in the 2500 light curve testing subset. The CKP is compared with the LS and AoV periodograms. The proposed method clearly outperforms its competitors in the FPR range of interest (below 1\%). It is worth noting that even if a harmonic of the true period is found, periodicity can still be detected. This is true for all the methods as periodicity detection rates are comparatively better than Hit rates obtained for period estimation.

\begin{table}[t]
	\begin{center}
	\caption{Period estimation performance of the CKP method versus other techniques for the testing database}  	 
	\begin{tabular}{l c c c}
	\hline

	\textbf{Method} & Hits[\%]& Multiples[\%]& Misses[\%] \\ \hline
	\textbf{CKP} & \textbf{88.00} & \textbf{11.60} & \textbf{0.4}\\
	CKP (linear kernel)  & $80.40$ & $18.67$ & $0.93$\\ 	
	Slotted correntropy & $78.80$ & $19.73$ & $1.47$ \\ 		
	Slotted correlation & $70.00$ & $28.80$ & $1.20$ \\
	VarTools LS & $61.73$ & $36.00$ & $2.27$ \\
	VarTools AoV & $75.33$ &  $23.60$ & $1.07$ \\ 	
	SLLK & $71.47$ & $26.27$ & $2.27$ \\
	\hline
	\end{tabular}
	\label{tab:Final}
	\end{center} \vspace{-10pt}
\end{table}

\section{Conclusion}

We have proposed a new metric for periodicity finding based on information theoretic concepts. The CKP metric yields a kernelized periodogram. It has been shown that the proposed method has several advantages over the correntropy spectral density and other conventional methods of period detection and estimation. The CKP is computed directly from the actual magnitudes and time-instants of samples. It does not require resampling, slotting nor folding schemes as other methods. The CKP metric is used as the main part of a fully-automated pipeline for period detection and estimation in astronomical time series. The CKP metric is used as test statistic to estimate the confidence level of the period detection.

Results on a subset of the MACHO survey shows that the CKP metric clearly outperforms the LS and AoV periodograms in the period detection of light curves. Moreover, the CKP method clearly outperforms slotted correntropy, slotted correlation, LS-periodogram, AoV and SLLK methods on the period estimation of periodic light curves. This is because the CKP method incorporates higher-order moments when computing the periodogram, and it emphasizes the trial period and its multiples. Future work will focus on enhancing the spurious rejection criterion, developing and adaptive kernel size adjusting rule, and discriminating quasi-periodic and multi-periodic light curves.

\section{Acknowledgement}

This work was funded by CONICYT-CHILE under grant FONDECYT 1110701, and its Doctorate Scholarship program. 

This paper utilizes public domain data obtained by the MACHO Project, jointly funded by the US Department of Energy through the University of California, Lawrence Livermore National Laboratory under contract No. W-7405-Eng-48, by the National Science Foundation through the Center for Particle Astrophysics of the University of California under cooperative agreement AST-8809616, and by the Mount Stromlo and Siding Spring Observatory, part of the Australian National University.

\bibliographystyle{hieeetr}
\addcontentsline{toc}{chapter}{Bibliography}
\bibliography{references}

\begin{thebibliography}{10}

\bibitem{Petit1997}
M.~Petit, {\em Variable Stars}.
\newblock Reading, MA: New York: Wiley, 1987.

\bibitem{Eyer1999}
L.~{Eyer}, ``{First Thoughts about Variable Star Analysis},'' {\em Baltic
  Astronomy}, vol.~8, pp.~321--324, 1999.

\bibitem{Debosscher2007}
J.~Debosscher, L.~Sarro, C.~Aerts, J.~Cuypers, B.~Vandenbussche, R.~Garrido,
  and E.~Solano, ``Automated supervised classification of variable stars.
  \text{I} methodology,'' {\em Astronomy \& Astrophysics}, vol.~475,
  pp.~1159--1183, 2007.

\bibitem{Wachman2009}
G.~Wachman, R.~Khardon, P.~Protopapas, and C.~Alcock, ``Kernels for periodic
  time series arising in astronomy,'' in {\em Proceedings of the European
  Conference on Machine Learning and Knowledge Discovery in Databases: Part
  II}, (Bled, Slovenia), pp.~489--505, 2009.

\bibitem{Popper1980}
D.~Popper, ``Stellar masses,'' {\em Annual review of astronomy and
  astrophysics}, vol.~18, pp.~115--164, 1980.

\bibitem{protopapas2005}
P.~{Protopapas}, R.~{Jimenez}, and C.~{Alcock}, ``{Fast identification of
  transits from light-curves},'' {\em Monthly Notices of the Royal Astronomical
  Society}, vol.~362, pp.~460--468, Sept. 2005, arXiv:astro-ph/0502301.

\bibitem{Alcock2000}
C.~Alcock {\em et~al.}, ``The \text{MACHO} project: Microlensing results from
  5.7 years of lmc observations,'' {\em The Astrophysical Journal}, vol.~542,
  pp.~281--307, 2000.

\bibitem{Udalski1997}
A.~Udalski, M.~Kubiak, and M.~Szymanski, ``Optical gravitational lensing
  experiment. \text{OGLE-II} -- the second phase of the \text{OGLE} project,''
  {\em Acta Astronomica}, vol.~47, pp.~319--344, 1997.

\bibitem{York2000}
D.~G. York {\em et~al.}, ``The sloan digital sky survey: Technical summary,''
  {\em The Astronomical Journal}, vol.~120, no.~3, p.~1579, 2000.

\bibitem{Kaiser2002}
N.~Kaiser {\em et~al.}, ``Pan-starrs: A large synoptic survey telescope
  array,'' {\em Society of Photo-Optical Instrumentation Engineers (SPIE)
  Conference Series}, vol.~4836, pp.~154--164, 2002.

\bibitem{LSST2012}
Z.~Ivezic {\em et~al.}, ``\text{LSST}: from science drivers to reference design
  and anticipated data products,'' June 2011.
\newblock Living document found at: http://www.lsst.org/lsst/overview/.

\bibitem{Principe2010}
J.~Principe, {\em Information Theoretic Learning: Renyi's Entropy and Kernel
  Perspectives}.
\newblock New York: Springer Verlag, 2010.

\bibitem{Lomb1976}
N.~Lomb, ``Least-squares frequency analysis of unequally spaced data,'' {\em
  Astrophysics and Space Science}, vol.~39, pp.~447--462, 1976.

\bibitem{Scargle1982}
J.~Scargle, ``Studies in astronomical time series analysis. ii. statistical
  aspects of spectral analysis of unevenly spaced data,'' {\em The
  Astrophysical Journal}, vol.~263, pp.~835--853, 1982.

\bibitem{Schwarzenberg1989}
A.~Schwarzenberg-Czerny, ``On the advantage of using analysis of variance for
  period search,'' {\em Monthly Notices of the Royal Astronomical Society},
  vol.~241, pp.~153--165, 1989.

\bibitem{Dworetsky1983}
M.~Dworetsky, ``A period finding method for sparse randonmly spaced
  observations,'' {\em Monthly Notices of the Royal Astronomical Society},
  vol.~203, pp.~917--923, 1983.

\bibitem{Clarke2002}
D.~Clarke, ``String/rope length methods using the lafler-kinman statistic,''
  {\em Astronomy \& Astrophysics}, vol.~386, pp.~763--774, 2002.

\bibitem{Mayo1974}
W.~Mayo, ``Spectrum measurements with laser velocimeters,'' in {\em Proceedings
  of dynamic flow conference, DISA Electronik A/S DK-2740}, (Skoolunder,
  Denmark), pp.~851--868, 1978.

\bibitem{Edelson1988}
R.~A. Edelson and J.~Krolik, ``The discrete correlation function: A new method
  for analyzing unevenly sampled variability data,'' {\em The Astrophysical
  Journal}, vol.~333, pp.~646--659, 1988.

\bibitem{Huijse2011}
P.~Huijse, P.~Estevez, P.~Zegers, J.~C. Principe, and P.~Protopapas, ``Period
  estimation in astronomical time series using slotted correntropy,'' {\em IEEE
  Signal Processing Letters}, vol.~18, no.~6, pp.~371--374, 2011.

\bibitem{Santamaria2006}
I.~Santamaria, P.~Pokharel, and J.~Principe, ``Generalized correlation
  function: Definition, properties, and application to blind equalization,''
  {\em IEEE Transactions on Signal Processing}, vol.~54, no.~6, pp.~2187--2197,
  2006.

\bibitem{Scholkopf2002}
B.~Sch\text{\"{o}}lkopf and A.~Smola, {\em Learning with Kernels}.
\newblock Cambridge, MA: Cambridge, MA: MIT Press, 2002.

\bibitem{Xu2008}
J.~Xu and J.~Principe, ``A pitch detector based on a generalized correlation
  function,'' {\em IEEE Transactions on Audio, Speech, and Language
  Processing}, vol.~16, no.~8, pp.~1420--1432, 2008.

\bibitem{Liu2007}
R.~Li, W.~Liu, and J.~C. Principe, ``A unifying criterion for instantaneous
  blind source separation based on correntropy,'' {\em IEEE Transactions on
  Signal Processing}, vol.~87, no.~8, pp.~1872--1881, 2007.

\bibitem{Gunduz2009}
A.~Gunduz and J.~C. Principe, ``Correntropy as a novel measure for nonlinearity
  tests,'' {\em IEEE Transactions on Signal Processing}, vol.~89, no.~1,
  pp.~14--23, 2009.

\bibitem{Michalak2010}
M.~Michalak, ``Time series prediction using periodic kernels,'' in {\em
  Computer Recognition Systems 4}, pp.~136--146, Berlin: Springer Verlag, 2010.

\bibitem{Rasmussen2006}
C.~E. Rasmussen and C.~K.~I. Williams, {\em Gaussian processes for machine
  learning}.
\newblock MIT Press, 2006.

\bibitem{Mackay1998}
D.~Mackay, {\em Introduction to Gaussian Processes}, vol.~168, pp.~133--165.
\newblock Springer, Berlin, 1998.

\bibitem{Wang2012}
Y.~Wang, R.~Khardon, and P.~Protopapas, ``{Nonparametric Bayesian Estimation of
  Periodic Functions},'' {\em ArXiv e-prints}, 2011, arXiv:abs/1111.1315v2.

\bibitem{Evans2000}
M.~Evans, N.~Hastings, and B.~Peacock, ``Von mises distribution,'' in {\em
  Statistical Distributions}, pp.~191--192, Wiley, 3~ed., 2000.

\bibitem{Abramovich2000}
F.Abramovich, T.~Bailey, and T.~Sapatinas, ``Wavelet analysis and its
  statistical applications,'' {\em The Statistician}, vol.~49 Part1, pp.~1--29,
  2000.

\bibitem{Jenkins1968}
G.~M. Jenkins and D.~G. Watts, {\em Spectral analysis and its applications}.
\newblock Holden-day, 1968.

\bibitem{Schmitz1999}
A.~Schmitz and T.~Schreiber, ``Testing for nonlinearity in unevenly sampled
  time series,'' {\em Phys. Rev. E}, vol.~59, no.~4, pp.~4044--4047, 1999.

\bibitem{Schreiber1999}
T.~Schreiber and A.~Schmitz, ``Surrogate time series,'' {\em Physica D:
  Nonlinear Phenomena}, vol.~142, pp.~346--382, 1999.

\bibitem{Buhlmann99}
P.~Buhlmann, ``Bootstraps for time series,'' {\em Statistical Sciense},
  vol.~17, no.~1, pp.~52--72, 1999.

\bibitem{Hartman2008}
J.~D. Hartman and et. al., ``Deep \text{MMT} transit survey of the open cluster
  \text{M37}. \text{II} variable stars,'' {\em The Astrophysical Journal},
  vol.~675, pp.~1254--1277, 2008.
\newblock Software available online at:
  http://www.cfa.harvard.edu/~jhartman/vartools/.

\bibitem{Devor2005}
J.~Devor, ``Solutions for 10,000 eclipsing binaries in the bulge fields of
  \text{OGLE II} using \text{DEBiL},'' {\em The Astrophysical Journal},
  vol.~628, pp.~411--425, 2005, arXiv:astro-ph/0504399.

\end{thebibliography}

\begin{IEEEbiography}[{\includegraphics[width=1in,height=1.25in,clip,keepaspectratio]{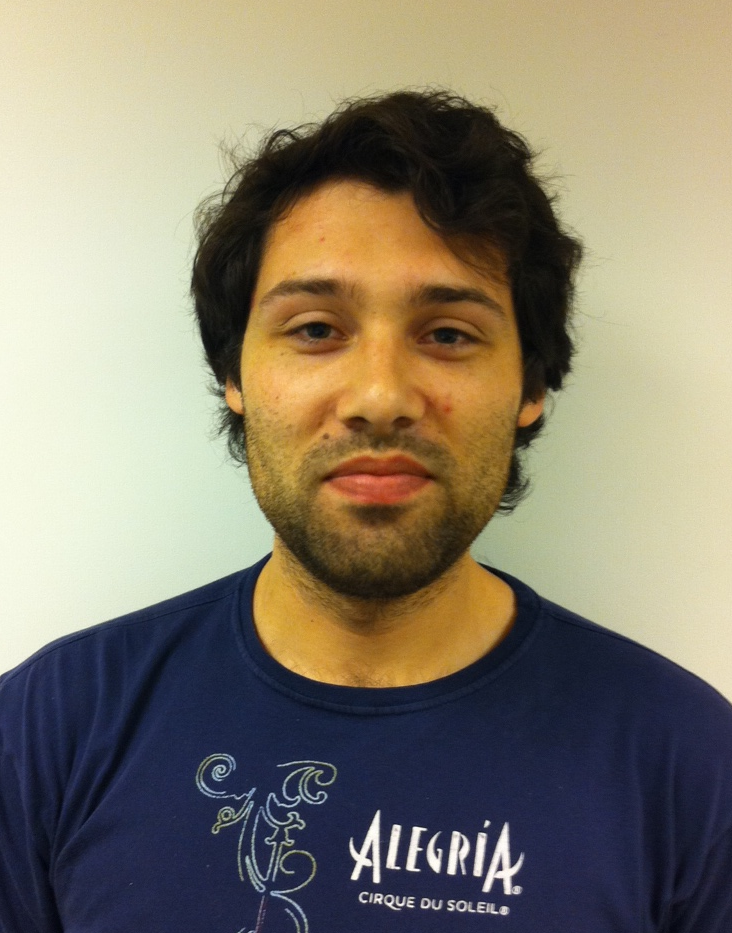}}]{Pablo Huijse} (GSM'10) was born in Valdivia, Chile in 1985. He received his B.S. and P.E. degrees in Electrical Engineering from the University of Chile in 2009. He is currently pursuing a PhD in Electrical Engineering at
the University of Chile. He was a visitor student at Harvard University in 2012. His research interests are primarily on astronomical time series analysis using information theoretic concepts.
\end{IEEEbiography}

\begin{IEEEbiography}[{\includegraphics[width=1in,height=1.25in,clip,keepaspectratio]{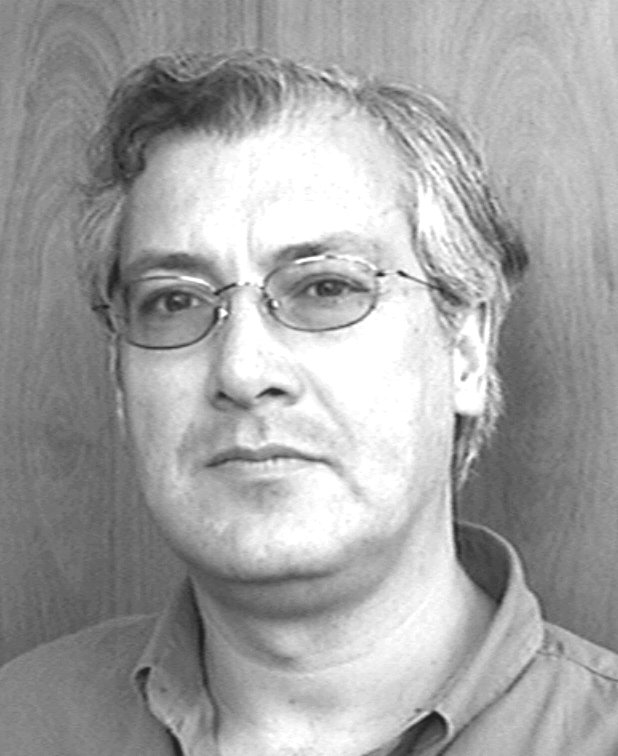}}]{Pablo Est\'evez} (M'98-SM'04) received the B.S. and the P.E. degrees in electrical engineering (EE) from the Universidad de Chile, Santiago, Chile, in 1978 and 1981, respectively, and the M. Eng. and Dr. Eng. degrees from the University of Tokyo, Japan, in 1992 and 1995, respectively. He is currently the Vice-Chairman and an Associate Professor of the EE Department, Universidad de Chile. He was an Invited Researcher at the NTT Communication Science Laboratory, Kyoto, Japan; the Ecole Normale Supérieure, Lyon, France, and a Visiting Professor at the University of Tokyo, Tokyo, Japan. Dr. Est\'evez is currently the Vicepresident for Members Activities of the IEEE Computational Intelligence Society (CIS). He has served as a Distinguished Lecturer and a member-at-large of the ADCOM of the IEEE CIS. He is an Associate Editor of the \textsc{IEEE Transactions on Neural Networks and Learning Systems}.
\end{IEEEbiography}

\begin{IEEEbiography}[{\includegraphics[width=1in,height=1.25in,clip,keepaspectratio]{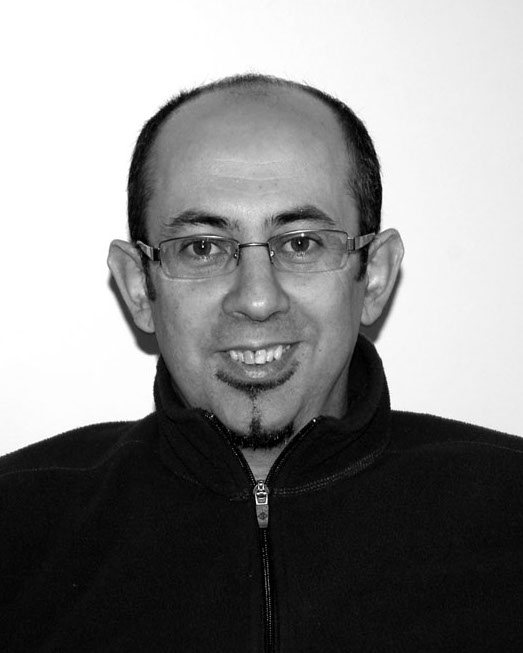}}]{Pavlos Protopapas} is a Lecturer at the School of Engineering and Applied Sciences at Harvard and a researcher at the Harvard Smithsonian Center for Astrophysics. He is the PI of the Time Series Center a multidisciplinary effort in the analysis of time series. He received his B.Sc. from Imperial College London and his Ph.D. from the University of Pennsylvania in theoretical physics. His research interests are primarily on time domain astronomy and large data analysis. 
\end{IEEEbiography}

\begin{IEEEbiography}[{\includegraphics[width=1in,height=1.25in,clip,keepaspectratio]{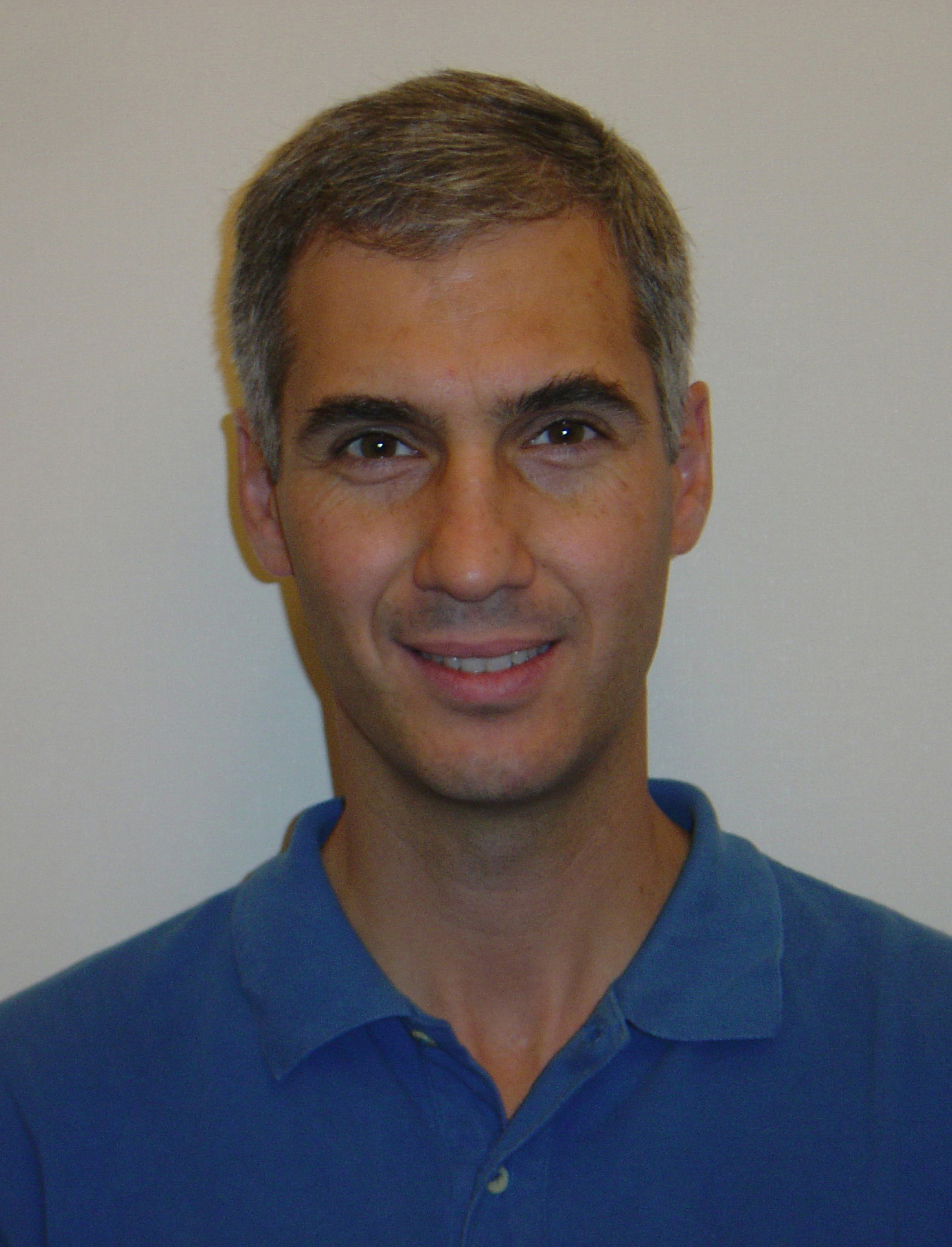}}]{Pablo Zegers} received his B.S. and P.E. degrees in Engineering from the Pontificia Universidad Cat\'olica, Chile, in 1992, his M.Sc. from The University of Arizona, USA, in 1998, and his Ph.D., also from The University of Arizona, in 2002. He is currently an Associate Professor of the College of Engineering and Applied Sciences of the Universidad de los Andes, Chile. His interests are artificial intelligence, machine learning, neural networks, and information theory. From 2006 to 2010 he was the Academic Director of this College, and for a brief period at the end of 2010, the Interim Dean. He is a Senior Member of the IEEE, and currently the Secretary of the Chilean IEEE section.
\end{IEEEbiography}

\begin{IEEEbiography}[{\includegraphics[width=1in,height=1.25in,clip,keepaspectratio]{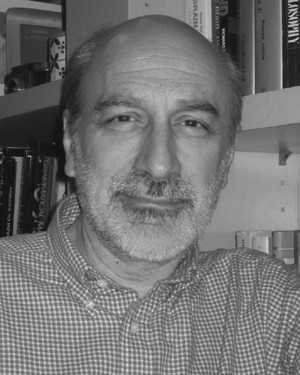}}]{Jos\'e Pr\'incipe} (M'83-SM'90-F'00) is currently a Distinguished Professor of Electrical and Biomedical Engineering at the University of Florida, Gainesville. He is BellSouth Professor and Founder and Director of the University of Florida Computational Neuro-Engineering Laboratory (CNEL). He is involved in biomedical signal processing, in particular, the  electroencephalogram (EEG) and the modeling and applications of adaptive systems. Dr. Pr\'incipe is the past Editor-in-Chief of the \textsc{IEEE Transactions on biomedical Engineering}, past President of the International Neural Network Society, former Secretary of the Technical Committee on Neural Networks of the IEEE Signal Processing Society, and a former member of the Scientific Board of the Food and Drug Administration. He was awarded the 2011 IEEE CIS Neural Networks Pioneer Award.
\end{IEEEbiography}

\end{document}